\documentclass[aps,prl,reprint,groupedaddress]{revtex4-2}
\usepackage[english]{babel}
\usepackage{graphicx} 
\usepackage{blindtext,bm,amsmath,amsfonts,physics,tikz,siunitx,mathtools,pgfplots,tabularx}
\definecolor{lila}{RGB}{103,70,235}
\usepackage{hyperref}
\hypersetup{
  colorlinks,
  citecolor=lila,
  linkcolor=lila,
  urlcolor=lila}

\usetikzlibrary{decorations.pathmorphing,decorations.pathreplacing,decorations.markings,arrows.meta,shapes.geometric,calc,fadings,shadings,bending}

\newcommand{\lo}[1]{_{#1}}
\newcommand{\hi}[1]{^{#1}}

\newcommand{\ud}{\mathrm{d}}

\renewcommand{\vec}[1]{\bm{#1}}
\newcommand{\mat}[1]{\mathrm{#1}}
\newcommand{\change}[1]{\textcolor{black}{#1}}
\newcommand{\pn}[1]{\hphantom{$-$}#1}
\newcommand{\nn}[1]{$-$#1}  
\definecolor{tegel}{RGB}{241, 90, 34} 
\definecolor{himmel}{RGB}{88, 176, 227}
\newcommand{\app}{appendix\xspace}
\newcommand{\appsection}{Appendix}

\begin{document}
\title{The phonomagnet: Spontaneous order of phonon angular
momentum}
\author{Maike Fahrensohn}
\affiliation{Department of Physics and Astronomy, Chalmers University of Technology, 412 96 G{\"o}teborg, Sweden}
\author{R. Matthias Geilhufe}
\affiliation{Department of Physics and Astronomy, Chalmers University of Technology, 412 96 G{\"o}teborg, Sweden}

\date{\today}

\begin{abstract}
Magnetism is a fundamental property of matter, typically arising from collective ordering of electronic spin and orbital angular momentum. Other quasiparticles interact with magnetism. However, they have not been considered its primary source. Here we show that phonons can generate a distinct form of magnetism through spontaneous ordering of their angular momentum. This phonomagnetism emerges from spin–phonon angular-momentum coupling, producing an effective exchange interaction between local ionic angular momenta. We show that the interaction strength depends on spin susceptibility, with a phonomagnetic phase predicted near a ferromagnetic critical point. Our theory also reveals phononic analogues of established spin models, including phonon Heisenberg, phonon $\Gamma$-model, and phonon Dzyaloshinskii–Moriya interactions. \change{Combining first-principles phonon calculations with recent data from Raman spectroscopy, we identify VI$_3$ as a candidate phonomagnet, providing a microscopic framework for interpreting the weak magnetic anomaly reported above its bulk ferromagnetic transition.} These findings open avenues for new platforms for information storage and manipulation on picosecond timescales.
\end{abstract}

\maketitle
Conventional magnetism arises from the collective ordering of electronic spin and orbital angular momentum. While phonons interact with electronic spins, they are rarely considered a source of magnetization. Chiral and axial phonons~\cite{Zhang2015ChiralLattices,Juraschek2025ChiralPhonons,Zhang2025ComprehensiveConstraints} are lattice excitations with a finite phonon angular momentum~\cite{McLellan1988AngularDynamics,zhang_angular_2014,Streib2021,Komiyama2021UniversalSymmetry,Bistoni2021IntrinsicMolecules,Saparov2022LatticePhonons,Coh2023,zhang_measurement_2025,Minakova2026ObservationModes}. It has been shown that the phonon angular momentum is closely connected to the formation of a phonon magnetic moment~\cite{rebane1983faraday,Juraschek2017DynamicalMultiferroicity,Juraschek2019OrbitalPhonons,Geilhufe2021DynamicallyKTaO_3,Ren2021PhononMagnetization,Juraschek2022GiantParamagnets,Geilhufe2023ElectronKaTaO_3,Fransson2023ChiralPolarization,Shabala2024PhononCoupling,Klebl2025UltrafastGeometry,Chaudhary2024GiantCoupling,Basini2024TerahertzSrTiO3,Merlin2024UnravelingFields,Merlin2024MagnetophononicsMisnomer,Shabala2025AxialEffects,Biggs2025UltrafastLiNbO3,Darmawan2026DisentanglingRT-TDDFT}, with pseudo-magnetic fields capable of controlling magnetization~\cite{Davies2024PhononicEffect}. 

Phonon angular momentum is discussed in three contexts. First, crystalline symmetry enforces chirality of selected phonon modes, as observed in 2D transition metal dichalcogenides~\cite{Zhang2015ChiralLattices,Zhu2018ObservationPhonons} and 3D chiral crystals~\cite{Ishito2023Truly-HgS,Ueda2023ChiralX-rays,Ishito2023ChiralTellurium,QiyangSun2026Momentum-resolvedTellurium}. Second, time-reversal symmetry breaking couples electronic magnetic ordering with phonon angular momentum, inducing circularly polarized phonon modes~\cite{Schaack1975Magnetic-fieldCeF3,Schaack1976ObservationField,Cheng2020ASemimetal,Baydin2022MagneticPbTe,Hernandez2023ObservationTopology,du_lattice_2019,Ren2024AdiabaticInsulators,Davies2024PhononicEffect,Fransson2025ChiralFerromagnetism,Bao2026Magnetic,Che2025MagneticSemimetal,Choi2026Magnetic}.  Third, driving IR-active phonon modes with circularly polarized THz lasers induces phonon angular momentum~\cite{Luo2023LargeHalides,Basini2024TerahertzSrTiO3,Davies2024PhononicEffect,Biggs2025UltrafastLiNbO3,Minakova2026ObservationModes}.
A fundamental question remains: Can phonon angular momentum order spontaneously, without external biases or pre-existing symmetry breaking? This would introduce a new type of magnetism, emerging entirely from ionic motion. To answer this question, we introduce a fourth mechanism: the spontaneous symmetry breaking of phonon angular momentum itself, mediated by an effective exchange interaction. 

\change{We show that this exchange interaction results from the hopping of itinerant electrons between lattice sites combined with the alignment of the electronic spin and ionic angular momentum. The latter emerges from the stochastic fluctuation of the ion around its equilibrium position. In general, this angular momentum averages to zero in thermal equilibrium, but its mean square does not. Like the velocity of a Brownian particle, the phonon angular momentum fluctuates about zero while its mean-square value is finite and grows with temperature. }

\change{We connect this mechanism to the layered magnet VI$_3$, where a weak magnetic anomaly precedes the bulk ferromagnetic transition~\cite{Gati2019,Valenta2021}. Using first-principles phonon calculations and experimentally reported mode splittings, we estimate the spin--phonon angular-momentum coupling and the resulting ordering temperature. This result connects the predicted phonomagnetic phase with existing experimental observations.}

\section*{From Spin-Phonon Coupling to Effective Phonon Interactions}
We begin with a Hamiltonian that captures the minimal ingredients for phonomagnetism,
\begin{equation}
H=H\lo{\text{ph}}+H\lo{\text{el}}+H\lo{\text{int}},        \label{eq:modelhamiltonian}
\end{equation}
where $H\lo{\text{ph}}$ and $H\lo{\text{el}}$ describe the phonons and electrons, respectively. $H\lo{\text{int}}$ is a \textit{spin-phonon angular momentum} (SPAM) coupling.

\begin{figure*}
    \centering
    \includegraphics[width=0.99\linewidth]{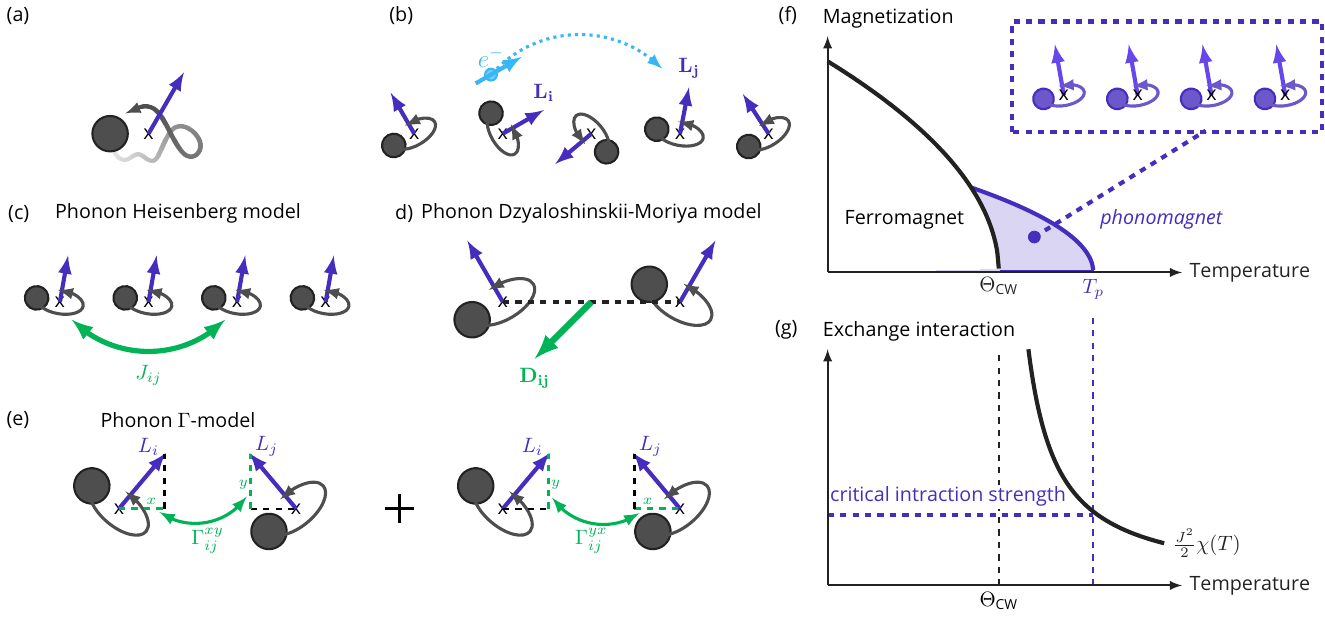}
    \caption{(a) The thermal fluctuations of ions close to their equilibrium positions induce a local ionic angular momentum. (b) Itinerant electrons on the lattice experience a weak spin polarization in the hopping process, giving rise to the spin-phonon angular momentum coupling. (c) In the absence of lattice frustration, the phonon Heisenberg model describes the collinear alignment of ionic angular momenta. (d) The phonon Dzyaloshinskii-Moriya interaction induces a canting of ionic angular momenta. (e) The phonon $\Gamma$-/pseudodipolar interaction acts as a source of bond-dependent frustration through off-diagonal symmetric interaction.  (f) Schematic phase diagram for a ferromagnet. Conventional spin order emerges below the Curie-Weiss temperature $\Theta_{\text{CW}}$. In contrast, the phonomagnetic phase is predicted to emerge at a higher temperature $T_p>\Theta_{\text{CW}}$. (g) The exchange interaction between ionic angular momenta is predicted to depend on the spin susceptibility. The divergence of the spin susceptibility at the Curie-Weiss temperature $\Theta_{\text{CW}}$ explains the realization of a critical interaction strength at $T_p>\Theta_{\text{CW}}$.}
    \label{fig:1}
\end{figure*}

The SPAM coupling arises from the Berry curvature of ionic motion. As an ion at site $i$ thermally fluctuates around its equilibrium position in a crystal (see \autoref{fig:1}(a)), its displacement $\vec{u}_i$ (including a factor of square root of mass) generates a non-Abelian Berry curvature \mbox{$\vec{\Omega}_{i}^{ab} = \mathrm{i}\, \nabla_{\vec{u}_{i}}\times\bra{a}\nabla_{\vec{u}_{i}}\ket{b}$}. This Berry curvature couples to the ionic angular momentum 
\mbox{$\vec{L}_{i} = \vec{u}_{i}\times\dot{\vec{u}}_{i}$}~\cite{zhang_angular_2014} as follows~\cite{shabala2026towards,Ren2021PhononMagnetization,Shabala2024PhononCoupling,Chaudhary2024GiantCoupling,Merlin2024UnravelingFields,Merlin2024MagnetophononicsMisnomer,Shabala2025AxialEffects,Klebl2025UltrafastGeometry,Darmawan2026DisentanglingRT-TDDFT}
\begin{equation}\label{eq:H_L_Berry_curvature}
    H_{\text{eff},i}^{ab} = \frac{\hbar}{2} \vec{L}_{i}\cdot \vec{\Omega}_{i}^{ab}.
\end{equation}
In a tight-binding picture, the states $\ket{a}$ and $\ket{b}$ can be local orbitals of ion $i$ contributing to a Bloch function of the crystal. According to equation~\eqref{eq:H_L_Berry_curvature}, the ionic angular momentum acts like a magnetic field, splitting electronic orbital multiplets. Hence, equation~\eqref{eq:H_L_Berry_curvature} corresponds to the coupling of the ionic angular momentum $\vec{L}_{i}$ and an electronic angular momentum of local orbitals, similar to the photonic counterpart discussed in Ref.~\cite{Pershan1966TheoreticalPhenomena}. For the sake of simplicity, we focus on either a pseudo-spin emerging from a two-fold degenerate orbital in the crystal field, or a real spin emerging from spin-orbit coupling, giving the SPAM interaction
\begin{equation}\label{Hint}
    H\lo{\text{int}}=-J\sum\limits\lo{i}\bm{L}\lo{i}(\tau)\cdot\bm{s}\lo{i}(\tau),
\end{equation} 
illustrated in \autoref{fig:1} (b). \change{$J$ is the effective coupling constant and the spin operator is given by \mbox{$\bm{s}\lo{i}=\frac{1}{2}\sum\lo{\alpha,\beta}c\lo{i\alpha}\hi{\dagger}(\tau)\sigma\hi{\alpha\beta}c\lo{i\beta}(\tau)$}. Here, we also allow for time-dependence in the electron spin and phonon angular momenta, denoted by the imaginary time $\tau$.}

Next, we use the SPAM coupling~\eqref{Hint} to derive an effective interaction between ionic angular momenta at two different sites $i$ and $j$. To do so, we formulate a partition function corresponding to the Hamiltonian~\eqref{eq:modelhamiltonian} and integrate out the electronic degrees of freedom. This allows us to formulate an effective action for the ionic degrees of freedom, including the effective interaction we are aiming for. The partition function of the entire system is given by
\begin{equation}
    Z(L)=\int\mathcal{D}u\mathcal{D}\Dot{u}\,e\hi{-S\lo{\text{ph}}(u\lo{i},\Dot{u}\lo{i})}Z\lo{\text{el}}(L),
    \label{eq:fullpartitionfunction}
\end{equation}
where $S\lo{\text{ph}}$ denotes the action corresponding to the phononic Hamiltonian, $H\lo{\text{ph}}$, and
\begin{equation}
    Z\lo{\text{el}}=Z\lo{0}\left\langle \mathcal{T}\lo{\tau}e\hi{J\int\limits\lo{0}\hi{\beta}\ud\tau \sum\limits\lo{i}\bm{L}\lo{i}(\tau)\bm{s}\lo{i}(\tau)}\right\rangle\lo{\text{el}}
    \label{eq:spinpartitionfunction}
\end{equation}
is the electron partition function in the presence of the SPAM interaction. $\mathcal{T}\lo{\tau}$ denotes imaginary time-ordering and $\langle\cdot\rangle\lo{\text{el}}$ denotes an average over electronic degrees of freedom. $Z\lo{0}$ is a normalization factor, given by
\begin{equation}
    Z\lo{0}\coloneqq\int\mathcal{D}c\hi{\dagger}\mathcal{D}c\, e\hi{-S\lo{\text{el}}(c\hi{\dagger},c)}.
\end{equation}
$S\lo{\text{el}}$ is the electron action corresponding to $H\lo{\text{el}}$. 

As we can see from equation~\eqref{eq:fullpartitionfunction}, the bare phononic part of the action is unaffected by the electronic integral. In contrast, electronic and SPAM interaction parts of the Hamiltonian are encoded in the electron partition function. Hence, the effective action for the phononic part is given by
\begin{equation}
    S\lo{\text{eff}}=S\lo{\text{ph}}-\ln Z\lo{\text{el}}.
\end{equation}

Expanding $\ln Z\lo{\text{el}}$ to second order in $J$ using a Dyson series, we obtain
\begin{multline}\label{eq:haminteraction}
    H\lo{\text{eff}}\hi{\text{int}}=-J\sum\limits\lo{i}\left\langle \vec{s}\lo{i}(\tau)\right\rangle\lo{\text{el}}\cdot\vec{L}\lo{i}(\tau)\\-\frac{J\hi{2}}{2}\int\ud\tau'\,\sum\limits\lo{i,j}\chi\lo{i\,j}\hi{\alpha\beta}(\tau-\tau')L\lo{i}\hi{\alpha}(\tau)L\lo{j}\hi{\beta}(\tau').
\end{multline}
Here, we use the spin susceptibility 
\begin{align}
    \chi\lo{i\,j}\hi{\alpha\beta}(\tau-\tau')&\coloneqq  \left\langle\mathcal{T}\lo{\tau}s\lo{i}\hi{\alpha}(\tau)s\lo{j}\hi{\beta}(\tau')\right\rangle\lo{\text{el}}\\&-\Big\langle s\lo{i}\hi{\alpha}(\tau)\Big\rangle\lo{\text{el}}\left\langle s\lo{j}\hi{\beta}(\tau')\right\rangle\lo{\text{el}}.  \nonumber
\end{align}
Equation~\eqref{eq:haminteraction} denotes the central result of our work. It describes the effective Hamiltonian of local ionic angular momenta. For a more detailed discussion of the derivation of the effective interaction, we refer to the \app.

The first term in equation~\eqref{eq:haminteraction} describes a Zeeman-like coupling between phonon angular momentum and electronic magnetization. This term breaks rotational symmetry, analogous to an external magnetic field in spin models such as the Ising or Heisenberg models. Physically, it describes the magnetization-induced circular polarization of phonons. Such a term gives rise to an intrinsic phonon Zeeman effect, splitting two-fold degenerate phonon lines into a left- and a right-circularly polarized mode~\cite{du_lattice_2019,Bao2026Magnetic,Che2025MagneticSemimetal,Choi2026Magnetic}. This phenomenon was observed, e.g., in the ferrimagnet Fe$_{2-x}$Zn$_x$Mo$_{3}$O$_{8}$, where magnetic order induces a giant spontaneous frequency splitting in the $E_{2}$ phonon modes, reaching as high as 25\% of the total phonon frequency~\cite{Choi2026Magnetic}. Similarly, in the kagome ferromagnetic Weyl material Co$_{3}$Sn$_{2}$S$_{2}$, spontaneous magnetic order induces a frequency splitting in the $E_{g}$ phonon mode~\cite{Che2025MagneticSemimetal}. Crucially, this bias term is present only in states with non-vanishing spin magnetization. In the absence of magnetic order, the linear interaction term vanishes and cannot facilitate phonon angular momentum ordering. 

The novel term in equation~\eqref{eq:haminteraction} is a bilinear exchange interaction between phonon angular momenta, mediated by the electronic spin susceptibility \mbox{$\chi\lo{i\,j}(\tau-\tau')$}. Unlike the linear term, this exchange does not require magnetization. It depends only on spin fluctuations, which persist even in paramagnetic states. Thus, it can drive spontaneous ordering of phonon angular momenta, enabling collective phononic phases (ferromagnetic, antiferromagnetic, or even topological) above the electronic transition temperature. We call this phenomenon \textit{phonomagnetism}.

\section*{Phononic Analogues of Spin Models}
The exchange term in equation~\eqref{eq:haminteraction} depends on the structure of the spin susceptibility $\chi\lo{ij}\hi{\alpha\beta}$. To resolve the physical meaning of this term, we decompose the spin susceptibility tensor into three distinct components according to its isotropy and symmetry properties:
\begin{multline}\label{symmetry}
    \frac{J\hi{2}}{2}\chi\lo{ij}\hi{\alpha\beta}(\tau-\tau') =\underbrace{J\lo{ij}(\tau-\tau')\delta\hi{\alpha\beta}}\lo{\text{isotropic, symmetric}}\\+\underbrace{\Gamma\lo{ij}\hi{\alpha\beta}(\tau-\tau')}\lo{\text{anisotropic, symmetric}} +\underbrace{\epsilon\hi{\alpha\beta\gamma}D\lo{ij}\hi{\gamma}(\tau-\tau')}\lo{\text{anisotropic, antisymmetric}}.
\end{multline}
The first tensor captures the isotropic and symmetric contributions, the second tensor the anisotropic and symmetric components, and the third tensor the anisotropic and antisymmetric part of the spin susceptibility. As a result, the  
exchange term in equation~\eqref{eq:haminteraction} maps onto phononic analogues of three canonical spin models: 
the Heisenberg~\cite{heisenberg_zur_1928,joyce_classical_1967}, $\Gamma$-/pseudodipolar~\cite{rau_generic_2014} and Dzyaloshinskii-Moriya model~\cite{moriya_anisotropic_1960,dzyaloshinsky_thermodynamic_1958} (see \autoref{fig:1}~(c)-(e)), 
\begin{align}
    H\lo{\text{eff}}\hi{\text{int}}=&-J\sum\limits\lo{i}\langle\bm{s}\lo{i}(\tau)\rangle\lo{\text{el}}\cdot\vec{L}\lo{i}(\tau)\nonumber\\&-\int\ud\tau'\underbrace{\sum\limits\lo{i,j}J\lo{ij}(\tau-\tau')\bm{L}\lo{i}(\tau)\cdot\bm{L}\lo{j}(\tau')}\lo{\text{phonon Heisenberg model}}\nonumber\\
    &-\int\ud\tau'\underbrace{\sum\limits\lo{i,j}\Gamma\lo{ij}\hi{\alpha\beta}(\tau-\tau')L\lo{i}\hi{\alpha}(\tau)L\lo{j}\hi{\beta}(\tau')}\lo{\text{phonon $\Gamma$-model}}\nonumber\\
    &-\int\ud\tau'\underbrace{\sum\limits\lo{i,j}\bm{D}\lo{ij}(\tau-\tau')\cdot\left(\bm{L}\lo{i}(\tau)\times\bm{L}\lo{j}(\tau')\right)}\lo{\text{phonon Dzyaloshinskii-Moriya interaction}}.\label{eq:hameffint}
\end{align}
The strength of these interactions is determined by the structure of the spin susceptibility tensor $\chi\lo{ij}\hi{\alpha\beta}$. This structure reflects the symmetries of the electronic Hamiltonian, with its anisotropy governed by spin-orbit coupling, crystal field effects, and spin correlations~\cite{gaiardoni_edelstein_2025,dos_santos_dias_anisotropic_2011}.

The phonon Heisenberg term is isotropic and contains only diagonal elements. Depending on the sign of $J\lo{ij}$ and in the absence of frustration, it stabilizes either parallel (ferromagnetic) or antiparallel (antiferromagnetic) alignment of the ionic angular momenta $\vec{L}_i$. The symmetric, off-diagonal anisotropic contribution to the bilinear term in equation~\eqref{eq:hameffint} corresponds to a phononic analogue of the $\Gamma$-model. Unlike the isotropic Heisenberg term, the $\Gamma$-model introduces direction-dependent coupling between phonon angular momenta, where the interaction strength varies with the relative orientation of $\vec{L}\lo{i}$ and $\vec{L}\lo{j}$. Similar to spin systems~\cite{Matsuda2025KitaevLiquids}, we expect this term to introduce strong frustration, stabilizing exotic phases such as chiral order or liquid behaviour of ionic angular momenta. Lastly, the anisotropic and antisymmetric contribution to the bilinear term in equation~\eqref{eq:hameffint} corresponds to a phononic analogue of the Dzyaloshinskii–Moriya interaction (DMI). By analogy with electronic DMI, we expect the phononic version to stabilize non-collinear and chiral textures in the spatial patterns of ionic angular momenta. In more complex scenarios, this could give rise to skyrmion-like phononic configurations, where $\vec{L}_i$ forms a topological pattern in real space.

\section*{Ordering of Phonon Angular Momentum Above the Ferromagnetic Phase Transition}
\change{
In the following we show the phonomagnetic phase transition in the vicinity of a ferromagnetic phase transition. For simplicity, we use the static approximation, with the static spin susceptibility
\begin{equation}
    \chi_{ij}^{\alpha\beta}
    = \int_0^{\beta} \ud\tau\,
    \chi_{ij}^{\alpha\beta}(\tau).
\end{equation}}
The spin susceptibility exhibits strong temperature dependence. In a ferromagnet, for example, approaching the Curie temperature from above, the increasing spin-spin correlation length enhances $\chi\lo{ij}\hi{\alpha\beta}$, following the Curie-Weiss law. Due to the divergence of the spin susceptibility near the Curie-Weiss temperature $\Theta\lo{\text{CW}}$, the effective phonon exchange interaction grows strong enough to drive ordering at a temperature \mbox{$T_p > \Theta\lo{\text{CW}}$}, i.e., above the electronic transition. We model this phenomenon with the phonon Heisenberg Hamiltonian
\begin{equation}\label{phonon_Heisenberg}
    H=H\lo{\text{ph}}-J\sum\limits\lo{i}\left\langle \vec{s}\lo{i}\right\rangle\lo{\text{el}}\cdot\bm{L}\lo{i}-\frac{1}{2}\sum\limits\lo{i,j} J\lo{ij}\bm{L}\lo{i}\bm{L}\lo{j},
\end{equation}
where $H\lo{\text{ph}}$ denotes the bare phonon Hamiltonian and \mbox{$J\lo{ij}=J\hi{2}\chi\lo{ij}$}. For simplicity, we assume static interactions. In mean-field theory, the Hamiltonian~\eqref{phonon_Heisenberg} simplifies to
\begin{equation}
    H\lo{\text{MF}}=H\lo{\text{ph}}-\sum\limits\lo{i}\bm{h}\lo{\text{eff},i}\cdot\bm{L}\lo{i},
\end{equation}
where $\bm{h}\lo{\text{eff},i} = J\left\langle \vec{s}\lo{i}\right\rangle\lo{\text{el}} + \sum_j J_{ij} \left<\vec{L}_j\right>$ is the local effective field. The expectation value of the ionic angular momentum $\left<\bm{L}\lo{i}\right>$ is proportional to the effective field $\bm{h}\lo{\text{eff},i}$,
\begin{equation}
    \left<\bm{L}\lo{i}\right> = \chi\lo{\text{L}} \bm{h}\lo{\text{eff},i},
\end{equation}
where $\chi\lo{\text{L}}$ is the phonon angular momentum susceptibility. The susceptibility $\chi\lo{\text{L}}$ is diagonal, as shown in the \app. 
In the absence of electronic magnetization ($\left\langle \vec{s}\lo{i}\right\rangle\lo{\text{el}} = 0$), the Fourier-transformed equation (\mbox{$\vec{L}_{\vec{q}} = \mathcal{F}\left[\left<\bm{L}\lo{i}\right>\right]$}) becomes
\begin{equation}\label{eigenvalueeq}
    \vec{L}\lo{\bm{q}}=J\hi{2}\chi\lo{\text{L}}(T)\chi(\bm{q},T)\vec{L}\lo{\bm{q}}.
\end{equation}
It is shown that $\chi\lo{\text{L}}$ increases linearly in temperature $T$ (see \app), i.e., \mbox{$\chi\lo{\text{L}} = 2 k_B T / \omega_0^2$}, with $k_B$ being the Boltzmann constant and $\omega_0$ the phonon frequency. Additionally, for a ferromagnet, the spin susceptibility follows the Curie-Weiss law, 
\begin{equation}
    \chi(\bm{q}=0,T)=\frac{C}{T-\Theta\lo{\text{CW}}},
\end{equation}
with $C$ being a material-specific constant. We solve equation~\eqref{eigenvalueeq} to obtain the transition temperature for the collective ordering of ionic angular momentum,
\begin{equation}
    T\lo{\text{p}}=\frac{\Theta\lo{\text{CW}}}{1-\frac{2k\lo{\text{B}}J\hi{2}C}{\omega\lo{0}\hi{2}}}>\Theta\lo{\text{CW}}.
    \label{eq:mfcrittemp}
\end{equation}
For a more detailed derivation of equation~\eqref{eq:mfcrittemp}, we refer to the \app. 
\change{At high temperatures $T\gg \Theta\lo{\text{CW}}$, the Curie law for a higher spin is given by $\chi = \frac{j(j+1)}{3 k_B T}$ (see \app), wich allows to approximate the Curie constant $C = \frac{j(j+1)}{3 k_B}$. The corresponding transition temperature into the phonomagnet follows from \eqref{eq:mfcrittemp},
\begin{equation}
    T\lo{\text{p}}=\frac{\Theta\lo{\text{CW}}}{1-\frac{2 j(j+1)J\hi{2}}{3\omega\lo{0}\hi{2}}}.
\end{equation}}
\change{Note that $J$ is typically much smaller than $\omega_0$. Hence, the denominator in equation~\eqref{eq:mfcrittemp} is a positive number $<1$, we deduce that \mbox{$T_p > \Theta_{\text{CW}}$}, as illustrated in \autoref{fig:1} (f)-(g). This is a central prediction of our work: phonon angular momentum can order above the electronic transition temperature. }
\begin{table}[b]
\centering
\caption{\change{Phonon frequencies at the $\Gamma$ point for VI$_3$ (R$\overline{3}$): Comparison of this work with experimentally measured modes reported in Ref.~\cite{Hovank2022}. Frequencies in THz.}}
\label{tab:phonon_diff_exp}
\renewcommand{\arraystretch}{1.2}
\begin{tabular}{lcc}
\hline
 & This Work & Exp.~\cite{Hovank2022} \\
\hline
$E_g^1$ & 1.32 & 1.44 \\
$A_g^1$ & 2.30 & 2.11 \\
$E_u^1$ & 2.39 & 2.41 \\
$E_g^3$ & 2.91 & 2.89 \\
$E_u^2$ & 3.23 & 3.26 \\
$A_g^3$ & 3.89 & 3.78 \\
\hline
\end{tabular}
\end{table}
\change{
\section{Phonomagnetic phase transition in VI$_3$}
The splitting of degenerate phonon modes in the ferromagnetic phase is seen in various compounds, such as Fe$_5$GeTe$_2$~\cite{Mekap2026}, Co$_3$Sn$_2$S$_2$~\cite{Che2025MagneticSemimetal}, Fe$_{2-x}$Zn$_x$Mo$_3$O$_8$~\cite{Choi2026Magnetic,Bao2026Magnetic}, and VI$_3$~\cite{Hovank2022}. For the example of VI$_3$, this data is used to estimate the coupling constant $J$ and the phonomagnetic transition temperature $T_p$. V$^{3+}$ has an electron configuration of [Ar]$3d^2$ and a resulting spin of $j=1$. We approximate the Curie-Weiss temperature to coincide with the transition temperature of the ferromagnetic phase transition. For VI$_3$ it is \mbox{$\Theta_{\text{CW}} = 49.5~\text{K}$}~\cite{Gati2019,Tian2019,Kong2019,Valenta2021}. Furthermore, VI$_3$ has a structural phase transition at 79~K, with a rhombohedral (R$\overline{3}$) high-temperature structure and a monoclinic (C2/m) low-temperature structure. At 36~K, VI$_3$ has a second magnetic-to-magnetic phase transition, with a further structural change to a triclinic phase. The VI$_3$ unit cell contains two formula units, i.e., 8 atoms. For the rhombohedral structure, the resulting 24 phonon branches exhibit the following symmetries at the $\Gamma$-point:  $4 A_g \oplus 4 A_u \oplus 4 E_g \oplus 4 E_u $, with the three acoustic branches transforming as $A_u \oplus E_u$. We confirm these modes in ab initio calculations, where we obtained the force constant matrix from total energy calculations in the framework of density functional theory and computed phonon mode frequencies and polarization vectors using phonopy~\cite{Togo2023,Togo2023a} (see \app for details). Comparison of our numerical values with the reported frequencies from Ref.~\cite{Hovank2022} is given in \autoref{tab:phonon_diff_exp}. 
Furthermore, Raman spectroscopy reported in Ref.~\cite{Hovank2022} shows the splitting of the $E_g^3$ mode (2.89~THz) over the structural phase transition and the magnetic phase transition. With the structural phase transition, the mode splits into two modes, $A_g^7$ and $A_g^8$, belonging to the C2/m crystalline symmetry. At 65~K, the splitting of the two modes is $0.023~$THz centered around an average frequency of 2.94~THz~\cite{Hovank2022}. At 55~K, i.e., 6~K above the ferromagnetic phase transition, the average frequency jumps to 2.97~THz and the difference increases to $\Delta\omega=0.05~$THz~\cite{Hovank2022}. With the magnetic phase transition, the splitting first decreases to $\Delta\omega=0.026~$THz at 45~K and then increases to $\Delta\omega=0.043~$THz at 35~K~\cite{Hovank2022}. This splitting is a superposition of a structural and a magnetic contribution. }

\change{We use the magnetic contribution to fit the coupling constant $J$ between electronic spin and ionic angular momentum. At low temperatures, we assume $\abs{J\sum\limits\lo{i}\langle\bm{s}_i\rangle\cdot\bm{L}\lo{i}} \gg \abs{\frac{1}{2}\sum\limits\lo{i,j} J\lo{ij}\bm{L}\lo{i}\bm{L}\lo{j}}$. We transform the local term $J \sum_{i}\langle\vec{s}_i \rangle\cdot \vec{L}_i$ to phonon normal modes $u_{\vec{k} \nu}$, with 
$\vec{k}$ denoting the wave vector and $\nu$ the phonon branch (see \app for details). 
The corresponding classical equation of motion for a phonon mode is given by
\begin{equation}
   \ddot{u}_{\vec{k}\nu} +\omega_\nu^2 u_{\vec{k}\nu} - 2 J \sum_{\nu'} \Lambda_{\nu\nu'}(\vec{k}) \dot{u}_{\vec{k}\nu'} = 0,
\end{equation}
where we neglect noise and damping terms for simplicity~\cite{Qiao2026,Qiao2026a}.
The spin-coupling matrix $\mat{\Lambda}$ is given by
\begin{equation}
    \Lambda_{\nu\nu'}(\vec{k}) = \sum_{\alpha lmn} \xi^*_{\nu,\alpha l}(\vec{k})\epsilon_{lmn}\xi_{\nu',\alpha m}(\vec{k})\langle s_{\alpha n}\rangle,
\end{equation}
where $\xi_{\nu,\alpha l}$ is the phonon polarization vector of branch $\nu$ and site $\alpha$. The indices $l,m,n$ run over Cartesian directions $x, y, z$ and $\epsilon_{lmn}$ is the anti-symmetric Levi-Civita tensor. As a consequence of the Wigner-Eckart theorem~\cite{gtpack1}, the $\Lambda$-matrix is block diagonal in the irreducible representations. The block corresponding to the four pairs of phonon branches transforming as the irreducible representation $E_g$ is given in \autoref{tab:lambdamatrix_Eg}. The full matrix is given in the \app.}

\begin{table}[t]
\centering
\caption{\change{$E_g$ block of the non-zero matrix elements $\Lambda\lo{\nu\nu'}$ at $\Gamma$, evaluated for $\langle\bm{s}\rangle=0.8\bm{\hat{z}}$ with $\bm{\hat{z}}$ along the crystallographic $c$ axis. By symmetry, the doubly-degenerate $E_g$ modes couple only among themselves. Subscripts 1 and 2 distinguish the two partners of each $E_g$ mode.}}
\label{tab:lambdamatrix_Eg}
\footnotesize
\begin{tabular}{l|*{8}{c}}
\hline
 & {$E_{g;1}^{1}$} & {$E_{g;2}^{1}$} & {$E_{g;1}^{2}$} & {$E_{g;2}^{2}$} & {$E_{g;1}^{3}$} & {$E_{g;2}^{3}$} & {$E_{g;1}^{4}$} & {$E_{g;2}^{4}$} \\[2pt]
\hline
$E_{g;1}^{1}$ & & \nn{0.002} & \nn{0.010} & \pn{0.007} & \nn{0.005} & \pn{0.004} & \pn{0.032} & \nn{0.024} \\[2pt]
$E_{g;2}^{1}$ & \pn{0.002} & & \pn{0.007} & \pn{0.010} & \pn{0.004} & \pn{0.005} & \nn{0.024} & \nn{0.032} \\[2pt]
$E_{g;1}^{2}$ & \pn{0.010} & \nn{0.007} & & \pn{0.068} & & \pn{0.036} & \nn{0.004} & \nn{0.220} \\[2pt]
$E_{g;2}^{2}$ & \nn{0.007} & \nn{0.010} & \nn{0.068} & & \nn{0.036} & & \pn{0.220} & \nn{0.004} \\[2pt]
$E_{g;1}^{3}$ & \pn{0.005} & \nn{0.004} & & \pn{0.036} & & \pn{0.019} & & \nn{0.116} \\[2pt]
$E_{g;2}^{3}$ & \nn{0.004} & \nn{0.005} & \nn{0.036} & & \nn{0.019} & & \pn{0.116} & \\[2pt]
$E_{g;1}^{4}$ & \nn{0.032} & \pn{0.024} & \pn{0.004} & \nn{0.220} & & \nn{0.116} & & \pn{0.710} \\[2pt]
$E_{g;2}^{4}$ & \pn{0.024} & \pn{0.032} & \pn{0.220} & \pn{0.004} & \pn{0.116} & & \nn{0.710} & \\[2pt]
\hline
\end{tabular}
\end{table}

\begin{figure}
    \centering
    \includegraphics[width=\linewidth]{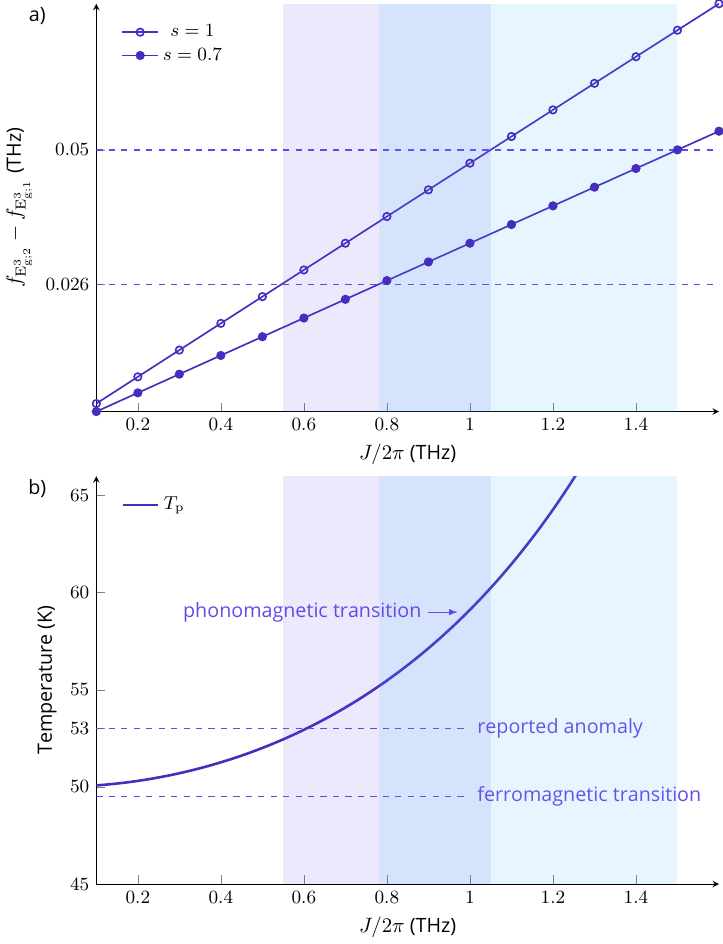}
    \caption{\change{Predicted phonomagnetic transition temperature in VI$_3$. a) Calculated splitting of the $E_g^3$ doublet using ab initio phonon frequencies and polarization vectors, for spin expectations $s=\vert\langle\bm{s}\rangle\vert=1$ (open circles) and $s=\vert\langle\bm{s}\rangle\vert=0.7$ (filled circles). Horizontal lines (dashed) delimit the splitting range, estimated from experimental data in Ref.~\cite{Hovank2022}. Shaded regions indicate corresponding intervals of the SPAM coupling strength $J$. b) Mean-field estimate of the phonomagnetic transition temperature, $T_{\mathrm{p}}$, as a function of $J$. The horizontal lines (dashed) mark the reported magnetic anomaly near $53\,\mathrm{K}$ and the ferromagnetic transition temperature at $49.5\,\mathrm{K}$~\cite{Gati2019,Valenta2021}. Shaded regions reproduce the coupling intervals from a).}}
    \label{fig:splitting}
\end{figure}

\change{
\autoref{fig:splitting} a) shows the inferred coupling for different spin contributions. The splitting of the $E_g^3$ doublet is calculated from ab initio phonon frequencies for different SPAM coupling constants $J$. Comparison with the experimental splitting range of approximately $0.026-0.05\,\mathrm{THz}$ yields a SPAM interaction \mbox{$J/2\pi\approx 0.55-1.05\,\mathrm{THz}$ for $\vert\langle\bm{s}\rangle\vert=1$} and \mbox{$J/2\pi\geq 1.2\,\mathrm{THz}$} for \mbox{$s=\vert\langle\bm{s}\rangle\vert=0.4$}. The estimate $s=0.4$ follows from a measured magnetic moment of approximately $0.8\mu_{\mathrm{B}}$ at around $50\,\mathrm{K}$~\cite{Valenta2021} when assuming a pure spin contribution with $g_\mathrm{s}=2$, while neglecting orbital magnetism. In contrast, an orbital magnetic moment of $0.6(1)\mu_\mathrm{B}$ was measured along the crystallographic $c$ axis in VI$_3$, antiparallel to the spin contribution~\cite{hovancik_large_2023}. Combining such an orbital magnetic contribution with a total moment of $0.8\mu_\mathrm{B}$ would imply a spin contribution of $\vert\langle \bm{s}\rangle\vert\approx 0.7$. The corresponding range for the SPAM coupling is then $J/(2\pi)\approx 0.78-1.50\,\mathrm{THz}$. This correction, however, does not establish the precise SPAM coupling range, as the orbital angular momentum was measured at $2\, \mathrm{K}$ and $5\,\mathrm{T}$, but it illustrates how orbital angular momentum can influence the determination of the SPAM coupling constant from comparison with experimental data. 
}
\change{From this approximation, the range of the critical temperature for the phonomagnetic transition can be estimated to $T_\mathrm{p}\approx 52.4-60.2\,\mathrm{K}$ for $\vert\langle\bm{s}\rangle\vert=1$ and $T_\mathrm{p}\approx 55.2-76.6\,\mathrm{K}$ for $\vert\langle\bm{s}\rangle\vert=0.7$.}

\change{
Interestingly, several recent experiments find a small anomaly in the magnetic susceptibility and magnetization at a temperature of 53~K, i.e., $\approx 3~\text{K}$ above the ferromagnetic phase transition~\cite{Valenta2021,Gati2019}. The sudden increase in magnetic moment per formula unit at 53~K is tiny, taking a value of $\approx 0.01~\mu_B$. This is consistent with the expected order of magnitude for an ionic contribution to magnetization due to chiral phonons~\cite{Juraschek2017DynamicalMultiferroicity,Juraschek2019OrbitalPhonons,Geilhufe2021DynamicallyKTaO_3}. In Ref.~\cite{Gati2019}, the origin of this small anomaly is assumed to stem from a surface contribution. In contrast, we propose the phonomagnetic phase as a potential explanation.}

\section*{Conclusion}
In summary, we have shown that the SPAM coupling between electron spin and phonon angular momentum in materials generates an effective exchange interaction between ionic angular momentum in materials. For sufficiently strong exchange interaction, this gives rise to the spontaneous symmetry breaking of the lattice and the formation of phonon angular momentum. Due to the time-reversal symmetry breaking of phonon angular momentum, spontaneous ordering of phonon angular momentum will be accompanied by a weak magnetic signal.

\change{The exchange interaction derived in this work is mediated by the electronic spin susceptibility. Through the spatial and temporal structure of the spin susceptibility, which acts as a retarded site-dependent coupling strength, this effective interaction inherits its isotropy and symmetry properties from the electronic interaction and maps onto phononic analogues of well-known spin interaction models such as the Heisenberg, $\Gamma-$, and Dzyaloshinskii-Moriya models. We demonstrate that this interaction can drive spontaneous symmetry breaking and support collective phonon ordering independent of long-range electronic order. As the phase transition is expected to occur above the Curie-Weiss temperature in magnetic materials, it corresponds to a distinct type of magnetic order, denoted as a phonomagnet.}

\change{Recent experiments on Fe$_5$GeTe$_2$~\cite{Mekap2026}, Co$_3$Sn$_2$S$_2$~\cite{Che2025MagneticSemimetal},
Fe$_{2-x}$Zn$_x$Mo$_3$O$_8$~\cite{Choi2026Magnetic,Bao2026Magnetic}, and VI$_3$~\cite{Hovank2022} demonstrate pronounced coupling between phonon dynamics and magnetism. Our framework identifies how this coupling can generate collective order of phonon angular momentum through electronic spin fluctuations, without requiring pre-existing
magnetic order. For VI$_3$, we combine ab initio phonon calculations with experimentally reported splittings of the $E_g^3$ doublet to estimate the SPAM coupling and the resulting phonomagnetic transition
temperature. These estimates support a phonomagnetic interpretation of the weak magnetic anomaly observed approximately 3~K above the bulk ferromagnetic transition. Refined polarization-resolved measurements
across this anomaly could test for the onset of phonon angular-momentum order, while measurements separating electronic spin and orbital moments would help constrain the inferred coupling strength. }

While we describe phonomagnetism classically, the framework admits a natural extension to the quantum mechanical regime through canonical quantization of the phonon modes. In this approach, one promotes the ionic displacement and momentum to bosonic operators in second quantization and defines the phonon angular momentum as the operator-valued cross product of displacement and velocity~\cite{zhang_angular_2014}. The spin-mediated phonon interaction can then be expressed as an effective interaction between quantized phonon modes, which would enable the study of genuinely quantum collective states of phonon angular momenta beyond the mean-field approximation. Near a magnetic quantum critical point, where the spin susceptibility becomes strongly enhanced or diverges under tuning of external parameters, the induced phonon interaction is likewise expected to become large. This raises the possibility of a phononic quantum phase transition at zero temperature.

We believe that this framework not only advances our understanding of spin-phonon angular momentum coupling but also opens up new possibilities for the study of collective phonon ordering phenomena, with potential implications for designing materials with novel magneto-phononic properties. More generally, the phonomagnet introduced in this work extends the current understanding of the origin of magnetism from electron spin and orbital angular momentum to other types of quasiparticles.

\bibliography{references}

@article{Juraschek2025ChiralPhonons,
    title = {{Chiral phonons}},
    year = {2025},
    journal = {Nature Physics},
    volume={21}, 
    pages={1532-1540},
    author = {Juraschek, Dominik M. and Geilhufe, R. Matthias and Zhu, Hanyu and Basini, Martina and Baum, Peter and Baydin, Andrey and Chaudhary, Swati and Fechner, Michael and Flebus, Benedetta and Grissonnanche, Gael and Kirilyuk, Andrei I. and Lemeshko, Mikhail and Maehrlein, Sebastian F. and Mignolet, Maxime and Murakami, Shuichi and Niu, Qian and Nowak, Ulrich and Romao, Carl P. and Rostami, Habib and Satoh, Takuya and Spaldin, Nicola A. and Ueda, Hiroki and Zhang, Lifa},
    month = {9},
    doi = {10.1038/s41567-025-03001-9},
    issn = {1745-2473}
}

@article{Zhu2018ObservationPhonons,
    title = {{Observation of chiral phonons}},
    year = {2018},
    journal = {Science},
    author = {Zhu, Hanyu and Yi, Jun and Li, Ming-Yang and Xiao, Jun and Zhang, Lifa and Yang, Chih-Wen and Kaindl, Robert A. and Li, Lain-Jong and Wang, Yuan and Zhang, Xiang},
    number = {6375},
    month = {2},
    pages = {579--582},
    volume = {359},
    doi = {10.1126/science.aar2711},
    issn = {0036-8075}
}

@article{Zhang2015ChiralLattices,
    title = {{Chiral Phonons at High-Symmetry Points in Monolayer Hexagonal Lattices}},
    year = {2015},
    journal = {Physical Review Letters},
    author = {Zhang, Lifa and Niu, Qian},
    number = {11},
    month = {9},
    pages = {115502},
    volume = {115},
    doi = {10.1103/PhysRevLett.115.115502},
    issn = {0031-9007}
}

@article{Zhang2025ComprehensiveConstraints,
    title = {{Comprehensive Study of Phonon Chirality under Symmetry Constraints}},
    year = {2025},
    journal = {arXiv:2503.22794},
    author = {Zhang, Shuai and Huang, Zhiheng and Du, Muchen and Ying, Tianping and Du, Luojun and Zhang, Tiantian},
    number = {  },
    month = {8},
    pages = {  },
    volume = {  },
    doi = {10.48550/arXiv.2503.22794},
    arxivId = {2503.22794}
}

@article{McLellan1988AngularDynamics,
    title = {{Angular momentum states for phonons and a rotationally invariant development of lattice dynamics}},
    year = {1988},
    journal = {Journal of Physics C: Solid State Physics},
    author = {McLellan, A G},
    number = {7},
    month = {3},
    pages = {1177--1190},
    volume = {21},
    doi = {10.1088/0022-3719/21/7/009},
    issn = {0022-3719}
}

@article{Coh2023,
  title = {Classification of materials with phonon angular momentum and microscopic origin of angular momentum},
  author = {Coh, Sinisa},
  journal = {Physical Review B},
  volume = {108},
  issue = {13},
  pages = {134307},
  numpages = {8},
  year = {2023},
  month = {Oct},
  publisher = {American Physical Society},
  doi = {10.1103/PhysRevB.108.134307},
  url = {https://link.aps.org/doi/10.1103/PhysRevB.108.134307}
}

@article{Streib2021,
  title = {Difference between angular momentum and pseudoangular momentum},
  author = {Streib, Simon},
  journal = {Physical Review B},
  volume = {103},
  issue = {10},
  pages = {L100409},
  numpages = {6},
  year = {2021},
  month = {Mar},
  publisher = {American Physical Society},
  doi = {10.1103/PhysRevB.103.L100409},
  url = {https://link.aps.org/doi/10.1103/PhysRevB.103.L100409}
}

@article{Minakova2026ObservationModes,
    title = {{Observation of angular momentum transfer among crystal lattice modes}},
    year = {2026},
    journal = {Nature Physics},
    author = {Minakova, Olga and Paiva, Carolina and Frenzel, Maximilian and Spencer, Michael S. and Urban, Joanna M. and Ringkamp, Christoph and Wolf, Martin and Mussler, Gregor and Juraschek, Dominik M. and Maehrlein, Sebastian F.},
    month = {5},
    volume={},
    pages={},
    doi = {10.1038/s41567-026-03274-8},
    issn = {1745-2473}
}

@article{Ishito2023Truly-HgS,
    title = {{Truly chiral phonons in {$\alpha$}-HgS}},
    year = {2023},
    journal = {Nature Physics},
    author = {Ishito, Kyosuke and Mao, Huiling and Kousaka, Yusuke and Togawa, Yoshihiko and Iwasaki, Satoshi and Zhang, Tiantian and Murakami, Shuichi and Kishine, Jun-ichiro and Satoh, Takuya},
    number = {1},
    month = {1},
    pages = {35--39},
    volume = {19},
    doi = {10.1038/s41567-022-01790-x},
    issn = {1745-2473}
}

@Article{rebane1983faraday,
  author  = {Rebane, Yu T},
  title   = {{Faraday} effect produced in the residual ray region by the magnetic moment of an optical phonon in an ionic crystal},
  journal = {Journal of Experimental and Theoretical Physics},
  year    = {1983},
  volume  = {84},
  pages   = {2323--2328},
}

@article{Juraschek2019OrbitalPhonons,
    title = {{Orbital magnetic moments of phonons}},
    year = {2019},
    journal = {Physical Review Materials},
    author = {Juraschek, Dominik M. and Spaldin, Nicola A.},
    number = {6},
    month = {6},
    pages = {064405},
    volume = {3},
    doi = {10.1103/PhysRevMaterials.3.064405},
    issn = {2475-9953}
}

@article{Juraschek2017DynamicalMultiferroicity,
    title = {{Dynamical multiferroicity}},
    year = {2017},
    journal = {Physical Review Materials},
    author = {Juraschek, Dominik M. and Fechner, Michael and Balatsky, Alexander V. and Spaldin, Nicola A.},
    number = {1},
    month = {6},
    pages = {014401},
    volume = {1},
    doi = {10.1103/PhysRevMaterials.1.014401},
    issn = {2475-9953}
}

@article{Juraschek2022GiantParamagnets,
    title = {{Giant effective magnetic fields from optically driven chiral phonons in $4f$ paramagnets}},
    year = {2022},
    journal = {Physical Review Research},
    author = {Juraschek, Dominik M. and Neuman, Tomáš and Narang, Prineha},
    number = {1},
    month = {2},
    pages = {013129},
    volume = {4},
    doi = {10.1103/PhysRevResearch.4.013129},
    issn = {2643-1564}
}

@article{Basini2024TerahertzSrTiO3,
    title = {{Terahertz electric-field-driven dynamical multiferroicity in SrTiO$_3$}},
    year = {2024},
    journal = {Nature},
    author = {Basini, M. and Pancaldi, M. and Wehinger, B. and Udina, M. and Unikandanunni, V. and Tadano, T. and Hoffmann, M. C. and Balatsky, A. V. and Bonetti, S.},
    number = {8008},
    month = {4},
    pages = {534--539},
    volume = {628},
    doi = {10.1038/s41586-024-07175-9},
    issn = {0028-0836}
}

@article{Davies2024PhononicEffect,
    title = {{Phononic switching of magnetization by the ultrafast Barnett effect}},
    year = {2024},
    journal = {Nature},
    author = {Davies, C. S. and Fennema, F. G. N. and Tsukamoto, A. and Razdolski, I. and Kimel, A. V. and Kirilyuk, A.},
    number = {8008},
    month = {4},
    pages = {540--544},
    volume = {628},
    doi = {10.1038/s41586-024-07200-x},
    issn = {0028-0836}
}

@article{Luo2023LargeHalides,
    title = {{Large effective magnetic fields from chiral phonons in rare-earth halides}},
    year = {2023},
    journal = {Science},
    author = {Luo, Jiaming and Lin, Tong and Zhang, Junjie and Chen, Xiaotong and Blackert, Elizabeth R. and Xu, Rui and Yakobson, Boris I. and Zhu, Hanyu},
    number = {6671},
    month = {11},
    pages = {698--702},
    volume = {382},
    doi = {10.1126/science.adi9601},
    issn = {0036-8075}
}

@article{Biggs2025UltrafastLiNbO3,
    title = {{Ultrafast Faraday Rotation Probe of Chiral Phonon-Polaritons in LiNbO$_3$}},
    year = {2025},
    journal = {arXiv: 2507.22232},
    author = {Biggs, Megan F. and {Sin-hang} and {Ho} and Alejandro, Aldair and Lutz, Matthew and Moss, Clayton D. and Johnson, Jeremy A.},
    number = { },
    month = {7},
    pages = { },
    volume = { },
    arxivId = {2507.22232},
    doi={10.48550/arXiv.2507.22232}
}

@article{Shabala2024PhononCoupling,
    title = {{Phonon Inverse Faraday Effect from Electron-Phonon Coupling}},
    year = {2024},
    journal = {Physical Review Letters},
    author = {Shabala, Natalia and Geilhufe, R. Matthias},
    number = {26},
    month = {12},
    pages = {266702},
    volume = {133},
    doi = {10.1103/PhysRevLett.133.266702},
    issn = {0031-9007}
}

@article{Shabala2025AxialEffects,
    title = {{Axial phono-magnetic effects}},
    year = {2025},
    journal = {arXiv:2511.03329},
    author = {Shabala, Natalia and Tietjen, Finja and Geilhufe, R. Matthias},
    number = { },
    month = {11},
    pages = { },
    volume = { },
    doi = {10.48550/arXiv.2511.03329},
    arxivId = {2511.03329}
}

@article{Geilhufe2023ElectronKaTaO_3,
    title = {{Electron magnetic moment of transient chiral phonons in KTaO$_3$}},
    year = {2023},
    journal = {Physical Review B},
    author = {Geilhufe, R. Matthias and Hergert, Wolfram},
    number = {2},
    month = {1},
    pages = {L020406},
    volume = {107},
    doi = {10.1103/PhysRevB.107.L020406},
    issn = {2469-9950}
}

@article{Geilhufe2021DynamicallyKTaO_3,
    title = {{Dynamically induced magnetism in KTaO$_3$}},
    year = {2021},
    journal = {Physical Review Research},
    author = {Geilhufe, R. Matthias and Juri{\v{c}}i{\'{c}}, Vladimir and Bonetti, Stefano and Zhu, Jian-Xin and Balatsky, Alexander V.},
    number = {2},
    month = {5},
    pages = {L022011},
    volume = {3},
    doi = {10.1103/PhysRevResearch.3.L022011},
    issn = {2643-1564}
}

@article{Ren2024AdiabaticInsulators,
    title = {{Adiabatic Dynamics of Coupled Spins and Phonons in Magnetic Insulators}},
    year = {2024},
    journal = {Physical Review X},
    author = {Ren, Shang and Bonini, John and Stengel, Massimiliano and Dreyer, Cyrus E. and Vanderbilt, David},
    number = {1},
    month = {3},
    pages = {011041},
    volume = {14},
    doi = {10.1103/PhysRevX.14.011041},
    issn = {2160-3308}
}

@article{Komiyama2021UniversalSymmetry,
    title = {{Universal features of canonical phonon angular momentum without time-reversal symmetry}},
    year = {2021},
    journal = {Physical Review B},
    author = {Komiyama, Hisayoshi and Murakami, Shuichi},
    number = {21},
    month = {6},
    pages = {214302},
    volume = {103},
    doi = {10.1103/PhysRevB.103.214302},
    issn = {2469-9950}
}

@article{Bistoni2021IntrinsicMolecules,
    title = {{Intrinsic Vibrational Angular Momentum from Nonadiabatic Effects in Noncollinear Magnetic Molecules}},
    year = {2021},
    journal = {Physical Review Letters},
    author = {Bistoni, Oliviero and Mauri, Francesco and Calandra, Matteo},
    number = {22},
    month = {6},
    pages = {225703},
    volume = {126},
    doi = {10.1103/PhysRevLett.126.225703},
    issn = {0031-9007}
}

@article{Saparov2022LatticePhonons,
    title = {{Lattice dynamics with molecular Berry curvature: Chiral optical phonons}},
    year = {2022},
    journal = {Physical Review B},
    author = {Saparov, Daniyar and Xiong, Bangguo and Ren, Yafei and Niu, Qian},
    number = {6},
    month = {2},
    pages = {064303},
    volume = {105},
    doi = {10.1103/PhysRevB.105.064303},
    issn = {2469-9950}
}

@article{Ren2021PhononMagnetization,
    title = {{Phonon Magnetic Moment from Electronic Topological Magnetization}},
    year = {2021},
    journal = {Physical Review Letters},
    author = {Ren, Yafei and Xiao, Cong and Saparov, Daniyar and Niu, Qian},
    number = {18},
    month = {10},
    pages = {186403},
    volume = {127},
    doi = {10.1103/PhysRevLett.127.186403},
    issn = {0031-9007}
}

@article{Merlin2024MagnetophononicsMisnomer,
    title = {{Magnetophononics and the chiral phonon misnomer}},
    year = {2024},
    journal = {PNAS Nexus},
    author = {Merlin, R},
    number = {1},
    month = {12},
    volume = {4},
    pages={},
    doi = {10.1093/pnasnexus/pgaf002},
    issn = {2752-6542}
}

@article{Merlin2024UnravelingFields,
    title = {{Unraveling the effect of circularly polarized light on reciprocal media: Breaking time reversal symmetry with non-Maxwellian magnetic-esque fields}},
    year = {2024},
    journal = {Physical Review B},
    author = {Merlin, R.},
    number = {9},
    month = {9},
    pages = {094312},
    volume = {110},
    doi = {10.1103/PhysRevB.110.094312},
    issn = {2469-9950}
}

@article{Fransson2023ChiralPolarization,
    title = {{Chiral phonon induced spin polarization}},
    year = {2023},
    journal = {Physical Review Research},
    author = {Fransson, J.},
    number = {2},
    month = {5},
    pages = {L022039},
    volume = {5},
    doi = {10.1103/PhysRevResearch.5.L022039},
    issn = {2643-1564}
}

@article{Fransson2025ChiralFerromagnetism,
    title = {{Chiral Phonons Enhance Ferromagnetism}},
    year = {2025},
    journal = {The Journal of Physical Chemistry Letters},
    author = {Fransson, Jonas and Kapon, Yael and Brann, Lilach and Yochelis, Shira and Sasselov, Dimitar D. and Paltiel, Yossi and Ozturk, S. Furkan},
    number = {8},
    month = {2},
    pages = {2001--2007},
    volume = {16},
    doi = {10.1021/acs.jpclett.5c00304},
    issn = {1948-7185}
}

@article{Bao2026Magnetic,
    title = {{Magnetic Signature of Chiral Phonons Revealed by Neutron Spectroscopy in Ferrimagnetic Fe$_{1.75}$Zn$_{0.25}$Mo$_3$O$_8$}},
    year = {2026},
    journal = {Physical Review Letters},
    author = {Bao, Song and Liao, Junbo and Huang, Zhentao and Shangguan, Yanyan and Ma, Zhen and Zhang, Bo and Cheng, Shufan and Xu, Hao and Song, Zihang and Dong, Shuai and Wu, Maofeng and Kajimoto, Ryoichi and Nakamura, Mitsutaka and Fennell, Tom and Khalyavin, Dmitry and Wen, Jinsheng},
    number = {9},
    month = {3},
    pages = {096502},
    volume = {136},
    doi = {10.1103/pq5m-32wj},
    issn = {0031-9007}
}

@article{Darmawan2026DisentanglingRT-TDDFT,
    title = {{Disentangling the dynamics of transient spin and orbital magnetization in SrTiO$_3$ via the inverse Faraday effect from RT-TDDFT}},
    year = {2026},
    journal = {arXiv:2602.20693},
    author = {Darmawan, Andri and Gruner, Markus E. and Pentcheva, Rossitza},
    number = { },
    month = {2},
    pages = { },
    volume = { },
    doi={10.48550/arXiv.2602.20693}
}

@article{du_lattice_2019,
	title = {Lattice {Dynamics}, {Phonon} {Chirality}, and {Spin}–{Phonon} {Coupling} in {2D} {Itinerant} {Ferromagnet} {Fe3GeTe2}},
	volume = {29},
    issn = {1616-3028},
	doi = {10.1002/adfm.201904734},
	number = {48},
	journal = {Advanced Functional Materials},
	author = {Du, Luojun and Tang, Jian and Zhao, Yanchong and Li, Xiaomei and Yang, Rong and Hu, Xuerong and Bai, Xueyin and Wang, Xiao and Watanabe, Kenji and Taniguchi, Takashi and Shi, Dongxia and Yu, Guoqiang and Bai, Xuedong and Hasan, Tawfique and Zhang, Guangyu and Sun, Zhipei},
	year = {2019},
	pages = {1904734},
}

@article{Che2025MagneticSemimetal,
    title = {{Magnetic Order Induced Chiral Phonons in a Ferromagnetic Weyl Semimetal}},
    year = {2025},
    journal = {Physical Review Letters},
    author = {Che, Mengqian and Liang, Jinxuan and Cui, Yunpeng and Li, Hao and Lu, Bingru and Sang, Wenbo and Li, Xiang and Dong, Xuebin and Zhao, Le and Zhang, Shuai and Sun, Tao and Jiang, Wanjun and Liu, Enke and Jin, Feng and Zhang, Tiantian and Yang, Luyi},
    number = {19},
    month = {5},
    pages = {196906},
    volume = {134},
    doi = {10.1103/PhysRevLett.134.196906},
    issn = {0031-9007}
}

@article{Chaudhary2024GiantCoupling,
    title = {{Giant effective magnetic moments of chiral phonons from orbit-lattice coupling}},
    year = {2024},
    journal = {Physical Review B},
    author = {Chaudhary, Swati and Juraschek, Dominik M. and Rodriguez-Vega, Martin and Fiete, Gregory A.},
    number = {9},
    month = {9},
    pages = {094401},
    volume = {110},
    doi = {10.1103/PhysRevB.110.094401},
    issn = {2469-9950}
}

@article{Choi2026Magnetic,
    title = {{Magnetic Control of Chiral Hybridized Phonon Magnetic Moments in Ferrimagnets Fe$_{2‐x}$Zn$_x$Mo$_3$O$_8$}},
    year = {2026},
    journal = {Advanced Functional Materials},
    author = {Choi, Youngsu and Wulferding, Dirk and Pawbake, Amit and Faugeras, Clément and Kuppusamy, Sivasakthi and Sankar, Raman and Son, Gichan and Kim, Jeongwoo and Choi, Kwang‐Yong},
    month = {5},
    doi = {10.1002/adfm.202512707},
    issn = {1616-301X},
    pages={e12707}
}

@article{joyce_classical_1967,
    title = {Classical {Heisenberg} {Model}},
    volume = {155},
    url = {https://link.aps.org/doi/10.1103/PhysRev.155.478},
    doi = {10.1103/PhysRev.155.478},
    number = {2},
    urldate = {2026-06-02},
    journal = {Physical Review},
    publisher = {American Physical Society},
    author = {Joyce, G. S.},
    month = mar,
    year = {1967},
    pages = {478--491},
}

@article{Ueda2023ChiralX-rays,
    title = {{Chiral phonons in quartz probed by X-rays}},
    year = {2023},
    journal = {Nature},
    author = {Ueda, Hiroki and Garc{\'{i}}a-Fern{\'{a}}ndez, Mirian and Agrestini, Stefano and Romao, Carl P. and van den Brink, Jeroen and Spaldin, Nicola A. and Zhou, Ke-Jin and Staub, Urs},
    number = {7967},
    month = {6},
    pages = {946--950},
    volume = {618},
    doi = {10.1038/s41586-023-06016-5},
    issn = {0028-0836}
}

@article{Ishito2023ChiralTellurium,
    title = {{Chiral phonons: circularly polarized Raman spectroscopy and ab initio calculations in a chiral crystal tellurium}},
    year = {2023},
    journal = {Chirality},
    author = {Ishito, Kyosuke and Mao, Huiling and Kobayashi, Kaya and Kousaka, Yusuke and Togawa, Yoshihiko and Kusunose, Hiroaki and Kishine, Jun‐ichiro and Satoh, Takuya},
    number = {6},
    month = {6},
    pages = {338--345},
    volume = {35},
    doi = {10.1002/chir.23544},
    issn = {0899-0042}
}

@article{QiyangSun2026Momentum-resolvedTellurium,
    title = {{Momentum-resolved direct observation of chiral phonons in elemental tellurium}},
    year = {2026},
    journal = {Physical Review Letters},
    author = {{Qiyang Sun} and {Songrui Hou} and {Ayman H. Said} and {Yanzhong Pei} and {Jie Ma} and {Wei Tian} and {Chen Li}},
    number = { },
    month = {5},
    pages = { },
    volume = { },
    doi = {10.1103/5cks-j4ht},
    issn = {0031-9007}
}

@article{Baydin2022MagneticPbTe,
    title = {{Magnetic Control of Soft Chiral Phonons in PbTe}},
    year = {2022},
    journal = {Physical Review Letters},
    author = {Baydin, Andrey and Hernandez, Felix G. G. and Rodriguez-Vega, Martin and Okazaki, Anderson K. and Tay, Fuyang and Noe, G. Timothy and Katayama, Ikufumi and Takeda, Jun and Nojiri, Hiroyuki and Rappl, Paulo H. O. and Abramof, Eduardo and Fiete, Gregory A. and Kono, Junichiro},
    number = {7},
    month = {2},
    pages = {075901},
    volume = {128},
    doi = {10.1103/PhysRevLett.128.075901},
    issn = {0031-9007}
}

@article{Hernandez2023ObservationTopology,
    title = {{Observation of interplay between phonon chirality and electronic band topology}},
    year = {2023},
    journal = {Science Advances},
    author = {Hernandez, Felix G. G. and Baydin, Andrey and Chaudhary, Swati and Tay, Fuyang and Katayama, Ikufumi and Takeda, Jun and Nojiri, Hiroyuki and Okazaki, Anderson K. and Rappl, Paulo H. O. and Abramof, Eduardo and Rodriguez-Vega, Martin and Fiete, Gregory A. and Kono, Junichiro},
    number = {50},
    month = {12},
    pages = {eadj407},
    volume = {9},
    doi = {10.1126/sciadv.adj4074},
    issn = {2375-2548}
}

@article{Schaack1975Magnetic-fieldCeF3,
    title = {{Magnetic-field dependent phonon states in paramagnetic CeF3}},
    year = {1975},
    journal = {Solid State Communications},
    author = {Schaack, G.},
    number = {4},
    month = {8},
    pages = {505--509},
    volume = {17},
    doi = {10.1016/0038-1098(75)90488-3},
    issn = {00381098}
}

@article{Schaack1976ObservationField,
    title = {{Observation of circularly polarized phonon states in an external magnetic field}},
    year = {1976},
    journal = {Journal of Physics C: Solid State Physics},
    author = {Schaack, G},
    number = {11},
    month = {6},
    pages = {L297-L301},
    volume = {9},
    doi = {10.1088/0022-3719/9/11/009},
    issn = {0022-3719}
}

@article{Cheng2020ASemimetal,
    title = {{A Large Effective Phonon Magnetic Moment in a Dirac Semimetal}},
    year = {2020},
    journal = {Nano Letters},
    author = {Cheng, Bing and Schumann, T. and Wang, Youcheng and Zhang, X. and Barbalas, D. and Stemmer, S. and Armitage, N. P.},
    number = {8},
    month = {8},
    pages = {5991--5996},
    volume = {20},
    doi = {10.1021/acs.nanolett.0c01983},
    issn = {1530-6984}
}

@article{Klebl2025UltrafastGeometry,
    title = {{Ultrafast Pseudomagnetic Fields from Electron-Nuclear Quantum Geometry}},
    year = {2025},
    journal = {Physical Review Letters},
    author = {Klebl, Lennart and Schobert, Arne and Eckstein, Martin and Sangiovanni, Giorgio and Balatsky, Alexander V. and Wehling, Tim O.},
    number = {1},
    month = {1},
    pages = {016705},
    volume = {134},
    doi = {10.1103/PhysRevLett.134.016705},
    issn = {0031-9007}
}

@article{shabala2026towards,
  title = {Towards unifying different microscopic theories for the phono-magnetic effect},
  author = {Shabala, Natalia and Liljegren, Ylva and Tietjen, Finja and Geilhufe, R. Matthias},
  journal = {in preparation},
  year = {},
  month={},
  pages={},
  number={},
  volume={}
}

@article{Pershan1966TheoreticalPhenomena,
    title = {{Theoretical Discussion of the Inverse Faraday Effect, Raman Scattering, and Related Phenomena}},
    year = {1966},
    journal = {Physical Review},
    author = {Pershan, P. S. and van der Ziel, J. P. and Malmstrom, L. D.},
    number = {2},
    month = {3},
    pages = {574--583},
    volume = {143},
    doi = {10.1103/PhysRev.143.574},
    issn = {0031-899X}
}

@article{Matsuda2025KitaevLiquids,
    title = {{Kitaev quantum spin liquids}},
    year = {2025},
    journal = {Reviews of Modern Physics},
    author = {Matsuda, Yuji and Shibauchi, Takasada and Kee, Hae-Young},
    number = {4},
    month = {12},
    pages = {045003},
    volume = {97},
    doi = {10.1103/3m4m-3v59},
    issn = {0034-6861}
}

@article{Mekap2026,
   author = {Smrutiranjan Mekap and Andrzej Ptok and Jyoti Saini and Changgu Lee and Subhasis Ghosh and Anushree Roy},
   doi = {10.1103/fylb-8xlx},
   issn = {2469-9950},
   issue = {21},
   journal = {Physical Review B},
   month = {6},
   pages = {214304},
   title = {Anisotropic electron-phonon coupling and chiral phonons in the van der Waals room temperature ferromagnet    {Fe$_5$GeTe$_2$}},
   volume = {113},
   year = {2026}
}

@article{Hovank2022,
   author = {Dávid Hovančík and Dalibor Repček and Fedir Borodavka and Christelle Kadlec and Karel Carva and Petr Doležal and Marie Kratochvílová and Petr Kužel and Stanislav Kamba and Vladimír Sechovský and Jiří Pospíšil},
   doi = {10.1021/acs.jpclett.2c02944},
   issn = {1948-7185},
   issue = {48},
   journal = {The Journal of Physical Chemistry Letters},
   month = {12},
   pages = {11095-11104},
   title = {Terahertz Magnetic and Lattice Excitations in van der Waals Ferromagnet {VI$_3$}},
   volume = {13},
   year = {2022}
}

@article{Tian2019,
   author = {Shangjie Tian and Jian-Feng Zhang and Chenghe Li and Tianping Ying and Shiyan Li and Xiao Zhang and Kai Liu and Hechang Lei},
   doi = {10.1021/jacs.8b13584},
   issn = {0002-7863},
   issue = {13},
   journal = {Journal of the American Chemical Society},
   month = {4},
   pages = {5326-5333},
   title = {Ferromagnetic van der Waals Crystal {VI$_3$}},
   volume = {141},
   year = {2019}
}

@article{Kong2019,
   author = {Tai Kong and Karoline Stolze and Erik I. Timmons and Jing Tao and Danrui Ni and Shu Guo and Zoë Yang and Ruslan Prozorov and Robert J. Cava},
   doi = {10.1002/adma.201808074},
   issn = {0935-9648},
   issue = {17},
   journal = {Advanced Materials},
   month = {4},
   title = {{VI$_3$} -- a New Layered Ferromagnetic Semiconductor},
   volume = {31},
   year = {2019},
   pages={}
}

@article{Valenta2021,
   author = {J. Valenta and M. Kratochvílová and M. Míšek and K. Carva and J. Kaštil and P. Doležal and P. Opletal and P. Čermák and P. Proschek and K. Uhlířová and J. Prchal and M. J. Coak and S. Son and J-G. Park and V. Sechovský},
   doi = {10.1103/PhysRevB.103.054424},
   issn = {2469-9950},
   issue = {5},
   journal = {Physical Review B},
   month = {2},
   pages = {054424},
   title = {Pressure-induced large increase of Curie temperature of the van der Waals ferromagnet {VI$_3$}},
   volume = {103},
   year = {2021}
}

@article{Gati2019,
   author = {Elena Gati and Yuji Inagaki and Tai Kong and Robert J. Cava and Yuji Furukawa and Paul C. Canfield and Sergey L. Bud'ko},
   doi = {10.1103/PhysRevB.100.094408},
   issn = {2469-9950},
   issue = {9},
   journal = {Physical Review B},
   month = {9},
   pages = {094408},
   title = {Multiple ferromagnetic transitions and structural distortion in the van der Waals ferromagnet {VI$_3$} at ambient and finite pressures},
   volume = {100},
   year = {2019}
}

@book{Schwabl2006,
   author = {Franz Schwabl},
   city = {Berlin, Heidelberg},
   doi = {10.1007/3-540-36217-7},
   isbn = {978-3-540-32343-3},
   publisher = {Springer Berlin Heidelberg},
   title = {Statistical Mechanics},
   year = {2006}
}

@article{Togo2023,
  title = {Implementation strategies in phonopy and phono3py},
  volume = {35},
  issn = {1361-648X},
  url = {http://dx.doi.org/10.1088/1361-648X/acd831},
  doi = {10.1088/1361-648x/acd831},
  number = {35},
  journal = {Journal of Physics: Condensed Matter},
  publisher = {IOP Publishing},
  author = {Togo, Atsushi and Chaput, Laurent and Tadano, Terumasa and Tanaka, Isao},
  year = {2023},
  month = June,
  pages = {353001},
}

@article{Togo2023a,
  title = {First-principles Phonon Calculations with Phonopy and Phono3py},
  volume = {92},
  issn = {1347-4073},
  url = {http://dx.doi.org/10.7566/JPSJ.92.012001},
  doi = {10.7566/jpsj.92.012001},
  number = {1},
  journal = {Journal of the Physical Society of Japan},
  publisher = {Physical Society of Japan},
  author = {Togo, Atsushi},
  year = {2023},
  month = Jan,
}

@article{Kresse1996a,
  author = {Kresse, G. and Furthmüller, J.},
  title = {Efficient iterative schemes for \textit{ab initio} total-energy calculations using a plane-wave basis set},
  journal = {Physical Review B},
  year = {1996},
  volume = {54},
  issn = {1095-3795},
  url = {http://dx.doi.org/10.1103/PhysRevB.54.11169},
  doi = {10.1103/physrevb.54.11169},
  number = {16},
  publisher = {American Physical Society (APS)},
  month = Oct,
  pages = {11169–11186},
}

@article{moriya_anisotropic_1960,
	title = {Anisotropic {Superexchange} {Interaction} and {Weak} {Ferromagnetism}},
	volume = {120},
	doi = {10.1103/PhysRev.120.91},
	number = {1},
	journal = {Physical Review},
	author = {Moriya, Tôru},
	year = {1960},
	pages = {91--98},
}

@article{gaiardoni_edelstein_2025,
	title = {Edelstein {Effect} in {Isotropic} and {Anisotropic} {Rashba} {Models}},
	volume = {10},
	copyright = {http://creativecommons.org/licenses/by/3.0/},
	issn = {2410-3896},
	url = {https://www.mdpi.com/2410-3896/10/1/15},
	doi = {10.3390/condmat10010015},
	number = {1},
	urldate = {2026-06-01},
	journal = {Condensed Matter},
	publisher = {Multidisciplinary Digital Publishing Institute},
	author = {Gaiardoni, Irene and Trama, Mattia and Maiellaro, Alfonso and Guarcello, Claudio and Romeo, Francesco and Citro, Roberta},
	month = mar,
	year = {2025},
	pages = {15},
}

@article{dos_santos_dias_anisotropic_2011,
	title = {Anisotropic spin-spin correlations in {Mn${}_{1}$/$X$(111) ($X=$ Pd, Pt, Ag, and Au)}},
	volume = {83},
	url = {https://link.aps.org/doi/10.1103/PhysRevB.83.054435},
	doi = {10.1103/PhysRevB.83.054435},
	number = {5},
	urldate = {2026-05-18},
	journal = {Physical Review B},
	publisher = {American Physical Society},
	author = {dos Santos Dias, M. and Staunton, J. B. and Deak, A. and Szunyogh, L.},
	month = feb,
	year = {2011},
	pages = {054435},
}

@article{zhang_measurement_2025,
	title = {Measurement of phonon angular momentum},
	volume = {21},
	issn = {1745-2473, 1745-2481},
	url = {https://www.nature.com/articles/s41567-025-02952-3},
	doi = {10.1038/s41567-025-02952-3},
	number = {9},
	urldate = {2026-02-17},
	journal = {Nature Physics},
	author = {Zhang, H. and Peshcherenko, N. and Yang, Fazhi and Ward, T. Z. and Raghuvanshi, P. and Lindsay, L. and Felser, Claudia and Zhang, Yang and Yan, J.-Q. and Miao, H.},
	month = sep,
	year = {2025},
	pages = {1387--1391},
}

@article{zhang_angular_2014,
	title = {Angular {Momentum} of {Phonons} and the {Einstein}–de {Haas} {Effect}},
	volume = {112},
	copyright = {http://link.aps.org/licenses/aps-default-license},
	issn = {0031-9007, 1079-7114},
	url = {https://link.aps.org/doi/10.1103/PhysRevLett.112.085503},
	doi = {10.1103/PhysRevLett.112.085503},
	number = {8},
	urldate = {2026-02-17},
	journal = {Physical Review Letters},
	author = {Zhang, Lifa and Niu, Qian},
	month = feb,
	year = {2014},
	pages = {085503},
}

@article{dzyaloshinsky_thermodynamic_1958,
	title = {A thermodynamic theory of “weak” ferromagnetism of antiferromagnetics},
	volume = {4},
	issn = {0022-3697},
	url = {https://www.sciencedirect.com/science/article/pii/0022369758900763},
	doi = {10.1016/0022-3697(58)90076-3},
	number = {4},
	urldate = {2026-05-05},
	journal = {Journal of Physics and Chemistry of Solids},
	author = {Dzyaloshinsky, I.},
	month = jan,
	year = {1958},
	pages = {241--255},
}

@article{rau_generic_2014,
  title = {Generic Spin Model for the Honeycomb Iridates beyond the Kitaev Limit},
  author = {Rau, Jeffrey G. and Lee, Eric Kin-Ho and Kee, Hae-Young},
  journal = {Physical Review Letters},
  volume = {112},
  issue = {7},
  pages = {077204},
  numpages = {5},
  year = {2014},
  month = {Feb},
  publisher = {American Physical Society},
  doi = {10.1103/PhysRevLett.112.077204},
  url = {https://link.aps.org/doi/10.1103/PhysRevLett.112.077204}
}

@article{heisenberg_zur_1928,
	title = {Zur {Theorie} des {Ferromagnetismus}},
	volume = {49},
	issn = {0044-3328},
	url = {https://doi.org/10.1007/BF01328601},
	doi = {10.1007/BF01328601},
	number = {9},
	urldate = {2026-04-30},
	journal = {Zeitschrift für Physik},
	author = {Heisenberg, W.},
	month = sep,
	year = {1928},
	pages = {619--636},
}

@article{Dudarev1998,
  author = {Dudarev, S. L. and Botton, G. A. and Savrasov, S. Y. and Humphreys, C. J. and Sutton, A. P.},
  title = {Electron-energy-loss spectra and the structural stability of nickel oxide: An {LSDA+U} study},
  journal = {Physical Review B},
  year = {1998},
  volume = {57},
  issn = {1095-3795},
  url = {http://dx.doi.org/10.1103/PhysRevB.57.1505},
  doi = {10.1103/physrevb.57.1505},
  number = {3},
  publisher = {American Physical Society (APS)},
  month = Jan,
  pages = {1505–1509},
}

@article{Qiao2026,
  title = {Quantum Heat Current in Terahertz‐Driven Phonon Systems},
  volume = {5},
  issn = {2751-1200},
  url = {http://dx.doi.org/10.1002/apxr.202500231},
  doi = {10.1002/apxr.202500231},
  number = {8},
  journal = {Advanced Physics Research},
  publisher = {Wiley},
  author = {Qiao, Yulong and Geilhufe, R. Matthias},
  year = {2026},
  month = June,
}

@misc{Qiao2026a,
  doi = {10.48550/ARXIV.2603.08027},
  url = {https://arxiv.org/abs/2603.08027},
  author = {Qiao, Fredrik Erikssonm Yulong and Fransson, Erik and Geilhufe, R. Matthias and Erhart, Paul},
  title = {Non-Markovian heat production in ultrafast phonon dynamics},
  publisher = {arXiv},
  year = {2026},
  copyright = {arXiv.org perpetual, non-exclusive license},
}

@article{Perdew1996,
  title = {Generalized Gradient Approximation Made Simple},
  volume = {77},
  issn = {1079-7114},
  url = {http://dx.doi.org/10.1103/PhysRevLett.77.3865},
  doi = {10.1103/physrevlett.77.3865},
  number = {18},
  journal = {Physical Review Letters},
  publisher = {American Physical Society (APS)},
  author = {Perdew, John P. and Burke, Kieron and Ernzerhof, Matthias},
  year = {1996},
  month = Oct,
  pages = {3865–3868},
}

@article{gtpack1,
Title = {{GTPack: A Mathematica Group Theory Package for Application in Solid-State Physics and Photonics}},
Author = {Geilhufe, R. Matthias and Hergert, Wolfram},
Journal = {Frontiers in Physics},
Year = {2018},
Pages = {86},
Volume = {6},
Doi = {10.3389/fphy.2018.00086},
ISSN = {2296-424X},
Url = {https://www.frontiersin.org/article/10.3389/fphy.2018.00086}
}

@article{hovancik_large_2023,
    title = {Large {Orbital} {Magnetic} {Moment} in {VI3}},
    volume = {23},
    issn = {1530-6984},
    url = {https://doi.org/10.1021/acs.nanolett.2c04045},
    doi = {10.1021/acs.nanolett.2c04045},
    number = {4},
    urldate = {2026-10-02},
    journal = {Nano Letters},
    author = {Hovančík, Dávid and Pospíšil, Jiří and Carva, Karel and Sechovský, Vladimír and Piamonteze, Cinthia},
    month = feb,
    year = {2023},
    pages = {1175--1180},
}

\section*{\appsection}
\subsection*{Exchange interaction between phonon angular momenta}

To calculate the effective interaction terms in the Hamiltonian that arise from the spin-phonon angular momentum coupling term, we integrate out the electronic spin degrees of freedom in the action (equation~\eqref{eq:fullpartitionfunction} in the main text) and obtain the effective action
\begin{equation}
    S\lo{\text{eff}}(\bm{L})=S\lo{\text{ph}}(\bm{L})-\ln Z\lo{\text{el}}.
    \label{eq:Seff}
\end{equation}
$S\lo{\text{ph}}$ describes the action corresponding to the phononic Hamiltonian, $H\lo{\text{ph}}$, and the electronic partition function is given by
\begin{equation}
    Z\lo{\text{el}}=Z\lo{0}\left\langle \mathcal{T}\lo{\tau}e\hi{J\int\limits\lo{0}\hi{\beta}\ud\tau \sum\limits\lo{i}\bm{L}\lo{i}(\tau)\bm{s}\lo{i}(\tau)}\right\rangle\lo{\text{el}}.
\end{equation}
Here, $\langle\cdot\rangle\lo{\text{el}}$ denotes an average over electronic degrees of
freedom and $Z\lo{0}$ is the normalization
\begin{equation}
    Z\lo{0}\coloneqq\int\mathcal{D}c\hi{\dagger}\mathcal{D}c\, e\hi{-S\lo{\text{el}}(c\hi{\dagger},c)},
\end{equation}
where $S\lo{\text{el}}$ is the corresponding electronic action
\begin{equation}
    S\lo{\text{el}}(c\hi{\dagger},c)=\int\limits\lo{0}\hi{\beta}\ud\tau\left(\sum\limits\lo{i,\sigma}c\lo{i\sigma}\hi{\dagger}(\tau)\partial\lo{\tau}c\lo{i\sigma}(\tau)+H\lo{\text{el}}(c\hi{\dagger},c)\right).
\end{equation}
To calculate the electronic partition function in equation~\eqref{eq:Seff}, we express it in terms of the normalization $Z\lo{0}$ and the electronic expectation value of the time-ordered exponential of the SPAM action. We then expand this expectation value to second order in the SPAM coupling constant $J$ using an imaginary-time Dyson series, and subsequently expand the logarithm to second order in $J$, yielding
\begin{align}
    \ln Z_{\text{el}} &= \ln Z_{0} + \ln \left\langle \mathcal{T}_{\tau} \exp\!\left[ J \! \int\limits_{0}^{\beta} \!\!\ud\tau \sum_{i} \bm{L}_{i}(\tau) \cdot \bm{s}_{i}(\tau) \right] \right\rangle_{\!\!\text{el}} \notag \\
    &= \ln Z_{0} + \ln \Big\{ \langle 1 \rangle_{\text{el}}
        + J \! \int \ud\tau \sum_{i} L_{i}^{\alpha}(\tau) \langle s_{i}^{\alpha}(\tau) \rangle_{\text{el}} \notag \\
    &\qquad + \frac{J^{2}}{2} \! \int \ud\tau \int \ud\tau' \sum_{i,j} L_{i}^{\alpha}(\tau) L_{j}^{\beta}(\tau') \notag\\
    &\qquad\times \langle \mathcal{T}_{\tau} s_{i}^{\alpha}(\tau) s_{j}^{\beta}(\tau') \rangle_{\text{el}} + \mathcal{O}(J^{3}) \Big\} \notag \\
    &= \ln Z_{0} + J \int \ud\tau \sum_{i} L_{i}^{\alpha}(\tau) \langle s_{i}^{\alpha}(\tau) \rangle_{\text{el}} \notag \\
    &\qquad + \frac{J^{2}}{2} \int \ud\tau \int \ud\tau' \sum_{i,j}
        L_{i}^{\alpha}(\tau) L_{j}^{\beta}(\tau') \notag \\
    &\qquad \times \Big[ \langle \mathcal{T}_{\tau} s_{i}^{\alpha}(\tau) s_{j}^{\beta}(\tau') \rangle_{\text{el}}
        - \langle s_{i}^{\alpha}(\tau) \rangle_{\text{el}} \langle s_{j}^{\beta}(\tau') \rangle_{\text{el}} \Big] \notag \\
    &= \ln Z_{0} + J \int \ud\tau \sum_{i} \langle s\lo{i}\hi{\alpha}(\tau)\rangle L_{i}^{\alpha}(\tau) \notag \\
    &\qquad + \frac{J^{2}}{2} \int \ud\tau \int \ud\tau' \sum_{i,j}
        \chi_{ij}^{\alpha\beta}(\tau - \tau') L_{i}^{\alpha}(\tau) L_{j}^{\beta}(\tau').
\end{align}
We define the spin susceptibility
\begin{multline}
    \chi\lo{ij}\hi{\alpha\beta}(\tau-\tau')\coloneqq\\ \left\langle\mathcal{T}\lo{\tau}s\lo{i}\hi{\alpha}(\tau)s\lo{j}\hi{\beta}(\tau')\right\rangle\lo{\text{el}}-\Big\langle s\lo{i}\hi{\alpha}(\tau)\Big\rangle\lo{\text{el}}\left\langle s\lo{j}\hi{\beta}(\tau')\right\rangle\lo{\text{el}}.
\end{multline}
Now, we can further split the second-order term into three parts,
\begin{align}
    \sum\limits\lo{i,j}\frac{J\hi{2}}{2}&\chi\lo{ij}\hi{\alpha\beta}(\tau-\tau')L\lo{i}\hi{\alpha}(\tau)L\lo{j}\hi{\beta}(\tau')\\
    &=\sum\limits\lo{i,j}\left\{J\lo{ij}(\tau-\tau')\bm{L}\lo{i}(\tau)\bm{L}\lo{j}(\tau')\right.\\
    &\qquad+\Gamma\lo{ij}\hi{\alpha\beta}(\tau-\tau')L\lo{i}\hi{\alpha}(\tau)L\lo{j}\hi{\beta}(\tau')\\
    &\qquad+\left.\bm{D}\lo{ij}(\tau-\tau')\left(\bm{L}\lo{i}(\tau)\times\bm{L}\lo{j}(\tau')\right)\right\}.
    \label{eq:splitsus}
\end{align}
Here, $\Gamma\lo{ij}\hi{\alpha\beta} = \Gamma\lo{ij}\hi{\beta\alpha}$ and $\Tr(\Gamma\lo{ij}) = 0$. The first term is isotropic and corresponds to a Heisenberg-like interaction. The anisotropic contributions split into a symmetric $\Gamma$-model/pseudodipolar term and an antisymmetric Dzyaloshinskii--Moriya term, represented by the second and third terms in equation~\eqref{eq:splitsus}, respectively.

\subsection*{Mean-field theory and phase transition}
Let us consider a simple example of an effective Hamiltonian as derived in equation~\eqref{eq:haminteraction},
\begin{equation}
    H=H\lo{\text{ph}}-J\sum\limits\lo{i}\left\langle \vec{s}\lo{i}\right\rangle\lo{\text{el}}\bm{L}\lo{i}-\frac{1}{2}\sum\limits\lo{i,j}J\lo{ij}\bm{L}\lo{i}\bm{L}\lo{j},
\end{equation}
consisting of a purely phononic part, a linear bias term, and a generalized Heisenberg-like bilinear phononic coupling. 

The phononic part of the Hamiltonian is given by the harmonic oscillator Hamiltonian for a free phonon
\begin{equation}
    H\lo{\text{ph}}=\sum\limits\lo{i}\frac{1}{2}\left(\Dot{\bm{u}}\lo{i}\hi{2}+\omega\lo{0}\hi{2}\bm{u}\lo{i}\hi{2}\right),
    \label{eq:hamapp}
\end{equation}
where we obtain the phonon modes from the phonon Langevin equations,
\begin{equation}
    \Ddot{\bm{u}}\lo{i}+\eta\Dot{\bm{u}}\lo{i}+\omega\lo{0}\hi{2}\bm{u}\lo{i}=\sqrt{2\eta k\lo{\text{B}}T}\xi(t),
\end{equation}
where $\eta$ is the damping coefficient, $\omega\lo{0}$ the natural frequency of the phonon mode, and $\xi(t)$ Gaussian white noise.

The second and third terms in equation~\eqref{eq:hamapp} correspond to the linear Zeeman-like effective contribution and the bilinear contribution, where we have restricted ourselves to the isotropic part of the interaction. We note that $J\lo{ij}=J\hi{2}\chi\lo{ij}$ depends on the spin phonon angular momentum coupling constant $J$ and the spin susceptibility $\chi\lo{ij}(\tau-\tau')$ of the spin background.  Here, we assume that the spin background is such that we can neglect time dependence and the spin susceptibility only has diagonal isotropic elements, i.e. $\chi\lo{ij}\hi{\alpha\beta}=\chi\lo{ij}\delta\hi{\alpha\beta}$ for sites $i,j$.

In the mean-field approximation, the Hamiltonian reduces to
\begin{equation}
    H\lo{\text{MF}}=H\lo{\text{ph}}-\sum\limits\lo{i}\bm{h}\lo{\text{eff},i}\bm{L}\lo{i},
\end{equation}
where $\bm{h}\lo{\text{eff},i}$ denotes the effective field $\bm{h}\lo{\text{eff},i}=J\left\langle \vec{s}\lo{i}\right\rangle\lo{\text{el}}+\sum\lo{j}J\lo{ij}\langle\bm{L}\lo{j}\rangle$ at site $i$. Here, $\mathcal{O}$ denotes the phonon expectation value of the phonon angular momentum defined by
\begin{equation}
    \langle\mathcal{O}\rangle\coloneqq\frac{1}{Z\lo{1}}\int\prod\limits\lo{j}\ud u\lo{j}\ud\Dot{u}\lo{j}\,\mathcal{O}\,e\hi{-\beta H\lo{\text{ph}}+\beta\sum\limits\lo{i}\bm{h}\lo{\text{eff},i}\bm{L}\lo{i}},
\end{equation}
with normalization
\begin{equation}
    Z\lo{1}(\bm{h}\lo{\text{eff}})\coloneqq \int\prod\limits\lo{j}\ud u\lo{j}\ud\Dot{u}\lo{j}\,e\hi{-\beta H\lo{\text{ph}}+\beta\sum\limits\lo{i}\bm{h}\lo{\text{eff},i}\bm{L}\lo{i}}.
\end{equation}
In linear response, the phonon expectation value can then be written as
\begin{equation}
    \langle L\lo
    i\hi{\alpha}\rangle=\sum\limits\lo{j,\beta}\chi\lo{\text{L},ij}\hi{\alpha\beta}h\lo{\text{eff},j}\hi{\beta}
    \label{eq:linearresponse}
\end{equation}
in terms of the phonon susceptibility tensor
\begin{equation}
    \chi\lo{\text{L},ij}\hi{\alpha\beta}=\pdv{\langle L\lo{i}\hi{\alpha}\rangle}{h\lo{\text{eff},j}\hi{\beta}}\Bigg\vert\lo{\bm{h}\lo{\text{eff}}=0}.
\end{equation}
It is important to note that here we denote the susceptibility tensor of the phonon angular momentum as $\chi\lo{L}$, in order not to confuse it with the spin susceptibility tensor $\chi$ that determines the coupling $J\lo{ij}=-J\hi{2}\chi\lo{ij}$ in the bilinear interaction term in the Hamiltonian.

The susceptibility can then be calculated to
\begin{multline}
    \chi\hi{\alpha\beta}\lo{\text{L},ij}(T)\\=\left.\pdv{}{h\lo{\text{eff},j}\hi{\beta}}\frac{1}{Z\lo{1}}\int\prod\limits\lo{l}\ud u\lo{l}\ud\Dot{u}\lo{l}\, L\lo{i}\hi{\alpha}\,e\hi{-\beta H\lo{\text{ph}}+ \beta \sum\limits\lo{k}\bm{h}\lo{\text{eff},k}\bm{L}\lo{k}}\right\vert\lo{\bm{h}\lo{\text{eff}}=0}\\
    =\beta\left(\langle L\lo{i}\hi{\alpha}L\lo{j}\hi{\beta}\rangle\lo{0}-\langle L\lo{i}\hi{\alpha}\rangle\lo{0}\langle L\lo{j}\hi{\beta}\rangle\lo{0}\right).
\end{multline}

Here, we use the notation
\begin{equation}
    \langle\mathcal{O}\rangle\lo{0}\coloneqq \langle\mathcal{O}\rangle\big\vert\lo{\bm{h}\lo{\text{eff}}=0}
\end{equation}
for the phonon expectation value at zero effective field $\bm{h}\lo{\text{eff}}$.

Calculating $\langle L\lo{i}\hi{\alpha}L\lo{j}\hi{\beta}\rangle\lo{0}-\langle L\lo{i}\hi{\alpha}\rangle\lo{0}\langle L\lo{j}\hi{\beta}\rangle\lo{0}$ in terms of the phonon modes $\bm{u}\lo{i}$ at site i and using Wick contraction, we see that
\begin{align}
    \chi\lo{\text{L},ij}\hi{\alpha\beta}&=\beta\epsilon\hi{\alpha\gamma\delta}\epsilon\hi{\beta\mu\nu}\Big(\langle u\lo{i}\hi{\gamma}\Dot{u}\lo{i}\hi{\delta}u\lo{j}\hi{\mu}\Dot{u}\lo{j}\hi{\nu}\rangle\lo{0}-\langle u\lo{i}\hi{\gamma}\Dot{u}\lo{i}\hi{\delta}\rangle\lo{0}\langle u\lo{j}\hi{\mu}\Dot{u}\lo{j}\hi{\nu}\rangle\lo{0}\Big)\\
    &=\beta\epsilon\hi{\alpha\gamma\delta}\epsilon\hi{\beta\mu\nu}\langle u\lo{i}\hi{\gamma}u\lo{j}\hi{\mu}\rangle\lo{0}\langle \Dot{u}\lo{i}\hi{\delta}\Dot{u}\lo{j}\hi{\nu}\rangle\lo{0}.
\end{align}
Here, the exponential in the integral in $\langle u\lo{i}\hi{\gamma}u\lo{j}\hi{\mu}\rangle\lo{0}$ factorizes into $\bm{u}$- and $\Dot{\bm{u}}$-dependent parts and $\langle u\lo{i}\hi{\gamma}\rangle\lo{0}=\langle\Dot{u}\lo{j}\hi{\mu}\rangle\lo{0}=0$ due to symmetry.

Due to the factorization of the integral in the free phonon expectation value, we can then solve the Gaussian integrals and obtain
\begin{equation}
    \langle u\lo{i}\hi{\gamma}u\lo{j}\hi{\mu}\rangle\lo{0}=\delta\lo{ij}\delta\hi{\gamma\mu}\frac{k\lo{\text{B}}T}{\omega\lo{0}\hi{2}},\quad \langle \Dot{u}\lo{i}\hi{\delta}\Dot{u}\lo{j}\hi{\nu}\rangle\lo{0}=\delta\lo{ij}\delta\hi{\delta\nu}k\lo{\text{B}}T.
\end{equation}

We see that the phonon susceptibility tensor is diagonal with elements
\begin{equation}
    \chi\lo{\text{L}}=\frac{2k\lo{\text{B}}T}{\omega\lo{0}\hi{2}}.
\end{equation}
Then, we can rewrite the linear response equation into an eigenvalue problem of the form
\begin{align}
    \langle L\lo{i}\hi{\alpha}\rangle&=\chi\lo{\text{L}}h\lo{\text{eff},i}\hi{\alpha}\\
    &=J\hi{2}\chi\lo{\text{L}}\sum\lo{j}\chi\lo{ij}\langle L\lo{j}\hi{\alpha}\rangle.
    \label{eq:selfconsistency}
\end{align}
For this to hold, $\langle L\hi{\alpha}\rangle$ must be an eigenvector of $\chi$ with eigenvalue $\mu$ such that
\begin{equation}
    1-J\hi{2}\chi\lo{\text{L}}\mu=0.
    \label{eq:evequation}
\end{equation}

To determine the critical temperature, we are looking for the eigenvalue $\mu$ that fulfills equation~\eqref{eq:evequation} at the highest temperature. Let us assume translational invariance for the spin background, i.e. $\chi\lo{ij}=\chi(\bm{R}\lo{i}-\bm{R}\lo{j})$. 
Starting from the self-consistency equation~\eqref{eq:selfconsistency}, we consider the Fourier transform in momentum space,
\begin{equation}
    m\lo{\bm{q}}\hi{\alpha}=J\hi{2}\chi\lo{\text{L}}(T)\chi(\bm{q},T)m\lo{\bm{q}}\hi{\alpha},
\end{equation}
where $\chi(\bm{q},T)$ is the Fourier transform of $\chi\lo{ij}$ and $m\lo{\bm{q}}\hi{\alpha}$ is the Fourier transform of $\langle L\lo{i}\hi{\alpha}\rangle$. 

For a ferromagnetic spin system, we can write the spin susceptibility in momentum space as
\begin{equation}\label{eq:spin_susc}
    \chi(\bm{q},T)=\frac{C}{T-\Theta\lo{\text{CW}}+Dq\hi{2}}.
\end{equation}
This is peaked at $\bm{q}=0$ and, thus, we obtain the critical temperature,
\begin{equation}
    T\lo{\text{p}}=\frac{\Theta\lo{\text{CW}}}{1-\frac{2k\lo{\text{B}}J\hi{2}C}{\omega\lo{0}\hi{2}}}.
\end{equation}

\change{
\section*{Spin susceptibility}
We motivate the Curie constant of the spin susceptibility \eqref{eq:spin_susc} from a high-temperature approximation. The Hamiltonian coupling the magnetic moment of a spin with an external magnetic field pointing along the $z$-direction is given by
\begin{equation}
    H = -\mu B_z\,s_z = -\frac{\mu B_z\,\sigma_z}{2}.
\end{equation}
Here, $\mu$ describes the amplitude of the magnetic moment. For example, for a single electron one has $\mu = \mu_B g_S$, with $\mu_B$ being the Bohr magneton and $g_S$ the Landé g-factor. It follows for the partition function,
\begin{equation}\label{eq:spin-part}
    Z = \operatorname{Tr} e^{-\beta H} = 2 \cosh{\left(\frac{\beta \mu B_z}{2}\right)}.
\end{equation}
Note that in~\eqref{Hint} we introduce the electron spin as a dimension-less operator, \mbox{$\bm{s}=\bm{S}/\hbar$}, which reduces to \mbox{$\bm{\sigma}/2$} in the spin-1/2 subspace. 
In this definition, the average spin follows from \eqref{eq:spin-part} as
\begin{equation}
    \langle s_z\rangle = \frac{1}{\beta\mu} \frac{\partial}{\partial B_z} \ln{Z} = \frac{1}{2}\tanh\left(\frac{\beta \mu B_z}{2}\right).
\end{equation}
At high temperatures, $k_B T \gg \vert\mu B_z\vert$ we can approximate $\langle s_z\rangle \approx \beta \mu B_z/4$. As a result, the spin susceptibility is given by
\begin{equation}\label{eq:spin_susc_th}
  \chi =  \frac{1}{\mu} \frac{\partial}{\partial B_z}\langle s_z\rangle \approx \frac{1}{4 k_B T}.
\end{equation}
Comparing equation \eqref{eq:spin_susc_th} with the Curie law $\chi = C/T$, we conclude $C = (4k_B)^{-1}$. 
In a similar manner, the spin susceptibility can be obtained for higher spins $j$~\cite{Schwabl2006}, leading to \begin{equation}\label{eq:spin_susc_higher}
  \chi = \frac{j(j+1)}{3 k_B T}.
\end{equation}
In this approximation, the transition temperature into the phonomagnet is then
\begin{equation}
    T\lo{\text{p}}=\frac{\Theta\lo{\text{CW}}}{1-\frac{2 j(j+1)J\hi{2}}{3\omega\lo{0}\hi{2}}}.
\end{equation}
}
\begin{table*}
\centering
\caption{\change{Non-zero matrix elements $\Lambda\lo{\nu\nu'}$ at the $\Gamma$-point, for $\vert\langle\bm{s}\rangle\vert=0.8$. Rows and columns are labeled by the corresponding phonon modes in the eigenvector ordering used in the calculation, following the notation of Table~\ref{tab:phonon_diff}. The first three modes are the acoustic branches, labeled Ac$_{i}$ ($i=1,\dots,3$); the two degenerate partners of each $E$ mode are distinguished by the subscripts 1 and 2.}}
\label{tab:lambdamatrix}
\renewcommand{\arraystretch}{1}
\setlength{\tabcolsep}{2pt}
\resizebox{\textwidth}{!}{
\begin{tabular}{l|*{24}{c}}
\hline
{Mode}
& {Ac$_1$} & {Ac$_2$} & {Ac$_3$} & {$E_{g;1}^{1}$} & {$E_{g;2}^{1}$} & {$A_u^{1}$}
& {$A_g^{1}$} & {$E_{g;1}^{2}$} & {$E_{g;2}^{2}$} & {$E_{u;1}^{1}$} & {$E_{u;2}^{1}$} & {$A_g^{2}$}
& {$E_{g;1}^{3}$} & {$E_{g;2}^{3}$} & {$E_{u;1}^{2}$} & {$E_{u;2}^{2}$} & {$A_g^{3}$} & {$A_u^{2}$}
& {$A_g^{4}$} & {$E_{u;1}^{3}$} & {$E_{u;2}^{3}$} & {$A_u^{3}$} & {$E_{g;1}^{4}$} & {$E_{g;2}^{4}$} \\[2pt]
\hline
{Ac$_1$} & & \pn{0.044} & \nn{0.078} & & & & & & & \pn{0.056} & \pn{0.018} & & & & \pn{0.008} & \nn{0.010} & & & & \pn{0.226} & \nn{0.071} & & &  \\[2pt]
{Ac$_2$} & \nn{0.044} & & \nn{0.031} & & & & & & & \pn{0.026} & \nn{0.024} & & & & \pn{0.008} & & & & & \pn{0.105} & \pn{0.098} & & &  \\[2pt]
{Ac$_3$} & \pn{0.078} & \pn{0.031} & & & & & & & & \nn{0.007} & \pn{0.054} & & & & \nn{0.008} & \nn{0.009} & & & & \nn{0.026} & \nn{0.219} & & &  \\[2pt]
$E_{g;1}^{1}$ & & & & & \nn{0.002} & & & \nn{0.010} & \pn{0.007} & & & & \nn{0.005} & \pn{0.004} & & & & & & & & & \pn{0.032} & \nn{0.024} \\[2pt]
$E_{g;2}^{1}$ & & & & \pn{0.002} & & & & \pn{0.007} & \pn{0.010} & & & & \pn{0.004} & \pn{0.005} & & & & & & & & &\nn{0.024} & \nn{0.032} \\[2pt]
$A_u^{1}$ & & & & & & & & & & & & & & & & & & & & & & & & \\
$A_g^{1}$ & & & & & & & & & & & & & & & & & & & & & & & & \\
$E_{g;1}^{2}$ & & & & \pn{0.010} & \nn{0.007} & & & & \pn{0.068} & & & & & \pn{0.036} & & & & & & & & & \nn{0.004} & \nn{0.220} \\[2pt]
$E_{g;2}^{2}$ & & & & \nn{0.007} & \nn{0.010} & & & \nn{0.068} & & & & & \nn{0.036} & & & & & & & & & & \pn{0.220} & \nn{0.004} \\[2pt]
$E_{u;1}^{1}$ & \nn{0.056} & \nn{0.026} & \pn{0.007} & & & & & & & & \nn{0.041} & & & & \pn{0.005} & \pn{0.007} & & & & \nn{0.001} & \pn{0.164} & & &  \\[2pt]
$E_{u;2}^{1}$ & \nn{0.018} & \pn{0.024} & \nn{0.054} & & & & & & & \pn{0.041} & & & & & \pn{0.007} & \nn{0.005} & & & & \pn{0.164} & \pn{0.001} & & &  \\[2pt]
$A_g^{2}$ & & & & & & & & & & & & & & & & & & & & & & & & \\
$E_{g;1}^{3}$ & & & & \pn{0.005} & \nn{0.004} & & & & \pn{0.036}  & & & & & \pn{0.019} & & & & & & & & & & \nn{0.116} \\[2pt]
$E_{g;2}^{3}$ & & & & \nn{0.004} & \nn{0.005} & & & \nn{0.036} & & & & & \nn{0.019} & & & & & & & & & & \pn{0.116} & \\[2pt]
$E_{u;1}^{2}$ & \nn{0.008} & \nn{0.008} & \pn{0.008} & & & & & & & \nn{0.005} & \nn{0.007} & & & & & \pn{0.002} & & & & \nn{0.021} & \pn{0.029} & & &  \\[2pt]
$E_{u;2}^{2}$ & \pn{0.010} & & \pn{0.009} & & & & & & & \nn{0.007} & \pn{0.005} & & & & \nn{0.002} & & & & & \nn{0.029} & \nn{0.021} & & &  \\[2pt]
$A_g^{3}$ & & & & & & & & & & & & & & & & & & & & & & & & \\
$A_u^{2}$ & & & & & & & & & & & & & & & & & & & & & & & & \\
$A_g^{4}$ & & & & & & & & & & & & & & & & & & & & & & & & \\
$E_{u;1}^{3}$ & \nn{0.226} & \nn{0.105} & \pn{0.026} & & & & & & & \pn{0.001} & \nn{0.164} & & & & \pn{0.021} & \pn{0.029} & & & & & \pn{0.663} & & &  \\[2pt]
$E_{u;2}^{3}$ & \pn{0.071} & \nn{0.098} & \pn{0.219} & & & & & & & \nn{0.164} & \nn{0.001} & & & & \nn{0.029} & \pn{0.021} & & & & \nn{0.663} & & & &  \\[2pt]
$A_u^{3}$ & & & & & & & & & & & & & & & & & & & & & & & & \\
$E_{g;1}^{4}$ & & & & \nn{0.032} & \pn{0.024} & & & \pn{0.004} & \nn{0.220} & & & & & \nn{0.116} & & & & & & & & & & \pn{0.710}  \\[2pt]
$E_{g;2}^{4}$ & & & & \pn{0.024} & \pn{0.032} & & & \pn{0.220} & \pn{0.004} & & & & \pn{0.116} & & & & & & & & & & \nn{0.710} &  \\[2pt]
\hline
\end{tabular}
}
\end{table*}
\change{
\section{Phonon equations of motion with background magnetization}
At low temperatures, we assume that the interaction with the background magnetization due to electronic spin dominates over the phonon-angular momentum exchange term, $J\sum\limits\lo{i}\langle\bm{s}_i\rangle\cdot\bm{L}\lo{i} \gg \frac{1}{2}\sum\limits\lo{i,j} J\lo{ij}\bm{L}\lo{i}\bm{L}\lo{j}$. In the following, we distinguish the sum over ions $i$ into a sum over cells $i$ and sites $\alpha$, $i\to (i,\alpha)$. The corresponding Lagrangian for ionic displacements is
\begin{equation}
    \mathcal{L} = \sum_{i\alpha} \left(\frac{\dot{\vec{u}}^2_{i\alpha}}{2}-\frac{1}{2}\sum_{j\beta} \vec{u}_{i\alpha}\cdot\mat{D}_{i\alpha,j\beta}\cdot\vec{u}_{j\beta}+J \langle \vec{s}_{i\alpha}\rangle\cdot\vec{L}_{i\alpha}\right). 
\end{equation}
Here, $\mat{D}_{i\alpha,j\beta}$ is the dynamical matrix. From the Lagrangian, we derive the corresponding equations of motion
\begin{equation}
    \ddot{\vec{u}}_{i\alpha} + \sum_{j\beta} \mat{D}_{i\alpha,j\beta}\cdot\vec{u}_{j\beta} - 2 J\, \dot{\vec{u}}_{i\alpha}\times\langle \vec{s}_{i\alpha}\rangle = 0.
\end{equation}
In the mean-field approximation of magnetic order, it follows $\langle\vec{s}_{i\alpha}\rangle \approx \langle\vec{s}_{\alpha}\rangle$. Assuming that the dynamical matrix is translation invariant, $\mat{D}_{i\alpha,j\beta} = \mat{D}_{\alpha,\beta}(\vec{R}_i-\vec{R}_j)$, we write the mass-weighted ionic displacement as $\vec{u}_{i\alpha} = e^{i \vec{k}\cdot\vec{R}_i} \vec{u}_{\vec{k}\alpha}$ and obtain
\begin{equation}
    \ddot{\vec{u}}_{\vec{k}\alpha} + \sum_{\beta} \mat{D}_{\alpha,\beta}(\vec{k})\cdot\vec{u}_{\vec{k}\beta} - 2 J\, \dot{\vec{u}}_{\vec{k}\alpha}\times\langle \vec{s}_{\alpha}\rangle = 0.
\end{equation}
Each vector $\vec{u}_{\vec{k}\alpha}$ has 3 Cartesian components ($x,y,z$). We transform from the $3N$-dimensional site-Cartesian basis to the phonon mode basis $\nu=1,\dots,3N$, by diagonalizing the dynamical matrix with eigenvectors
\begin{equation}
    \sum_{\beta m} D_{\alpha l,\beta m}(\vec{k})\cdot \xi_{\nu,\beta m}(\vec{k}) = \omega_\nu^2(\vec{k})\, \xi_{\nu,\alpha l}(\vec{k}),
\end{equation}
where we introduced the Cartesian index $l,m = x,y,z.$ Consequently, we transform the ionic displacements into phonon modes via
\begin{equation}
    u_{\vec{k}\alpha l} = \sum_{\nu} \xi_{\nu, \alpha l}(\vec{k}) u_{\vec{k}\nu}, \qquad u_{\vec{k}\nu} = \sum_{\alpha l} \xi^*_{\nu, \alpha l}(\vec{k}) u_{\vec{k}\alpha l}.
\end{equation}
Using the orthonormality $\sum_{\alpha l} \xi^*_{\nu,\alpha l}\xi_{\nu',\alpha l} = \delta_{\nu\nu'}$, we obtain
\begin{equation}\label{eq:mode-dynamics}
   \ddot{u}_{\vec{k}\nu} +\omega_\nu^2 u_{\vec{k}\nu} - 2 J \sum_{\nu'} \Lambda_{\nu\nu'}(\vec{k}) \dot{u}_{\vec{k}\nu'} = 0.
\end{equation}
The spin-coupling matrix $\mat{\Lambda}$ is given by
\begin{equation}
    \Lambda_{\nu\nu'}(\vec{k}) = \sum_{\alpha lmn} \xi^*_{\nu,\alpha l}(\vec{k})\epsilon_{lmn}\xi_{\nu',\alpha m}(\vec{k})\langle s_{\alpha n}\rangle,
\end{equation}
with $\epsilon_{lmn}$ being the anti-symmetric Levi-Civita tensor.
}

\begin{table}[b]
\centering
\caption{\change{Phonon frequencies at the $\Gamma$ point for VI$_3$ (R$\overline{3}$): Comparison of this work with recent experimental and theoretical work reported in Ref.~\cite{Hovank2022}. Frequencies in THz.}}
\label{tab:phonon_diff}
\renewcommand{\arraystretch}{1.2}
\begin{tabular}{lcccc}
\hline
\textbf{} & This Work & Exp.~\cite{Hovank2022} & Theo.~\cite{Hovank2022} & Activity \\ 
\hline
$E_g^1$ & 1.32 & 1.44 & 1.41 & Raman \\ 
$A_u^1$ & 1.93 &  & 2.12 & IR \\ 
$A_g^1$ & 2.30 & 2.11 & 2.24 & Raman \\ 
$E_u^1$ & 2.39 & 2.41 & 2.50 & IR \\ 
$E_g^2$ & 2.31 &  & 2.73 & Raman \\ 
$A_g^2$ & 2.57 &  & 2.60 & Raman \\ 
$E_g^3$ & 2.91 & 2.89 & 2.96 & Raman \\ 
$E_u^2$ & 3.23 & 3.26 & 3.43 & IR \\ 
$A_g^3$ & 3.89 & 3.78 & 3.81 & Raman \\ 
$A_u^2$ & 3.89 &  & 3.94 & IR \\ 
$A_g^4$ & 5.67 & & 5.93 & Raman \\ 
$E_u^3$ & 6.22 & & 6.79 & IR \\ 
$A_u^3$ & 7.74 & & 7.64 & IR \\ 
$E_g^4$ & 9.89 & & 6.98 & Raman \\ 
\hline
\end{tabular}
\end{table}

\change{
\section{Splitting of phonon frequency due to background magnetization in VI$_3$}
To estimate the coupling constant $J$, we compute the spin-coupling matrix $\Lambda$ for the van-der-Waals ferromagnet VI$_3$. We calculate the phonon eigenvectors $\xi_{\nu,\alpha l}$ from ab initio calculations in the harmonic approximation using phonopy~\cite{Togo2023,Togo2023a}. The force constant matrix is computed from total energy calculations in the framework of density functional theory using VASP\cite{Kresse1996a}.
We use the R$\overline{3}$ crystal structure for VI$_3$, obtained at 60~K in Ref.~\cite{Tian2019}. We perform calculations with an energy cut-off of 500~eV and a $12\times12\times12$ $\vec{k}$-point mesh. We use the generalized gradient approximation, according to Perdew, Burke, and Ernzerhof~\cite{Perdew1996}. Similar to Ref.~\cite{Hovank2022}, we correct the $d$-electrons of vanadium with a DFT+U scheme~\cite{Dudarev1998}, using a value of $U=3.8$. All ionic positions are relaxed to the computational ground state, while the cell volume and shape are kept fixed. We focus on the $\Gamma$-point phonons, where calculations on the conventional unit cell are sufficient. The frequencies of the computed phonon modes are listed in \autoref{tab:phonon_diff} and compared to previously reported values from Ref.~\cite{Hovank2022}. Our computed phonon frequencies are in good agreement with the experimental values, showing relative differences of less than 10~\%. Compared to previously reported theoretical values, we find an overall good agreement, with a few substantial differences, in particular for the modes $E_g^2$, $E_g^4$. The computed $\Lambda$-matrix is given in \autoref{tab:lambdamatrix}. Evaluating equation \eqref{eq:mode-dynamics}, we determine the splitting of the Raman mode $E_g^3$. }

\section*{acknowledgments}
We acknowledge helpful discussions with Natalia Shabala and Yulong Qiao.

\section*{Author Contributions}
M.F. and R.M.G. developed the theoretical framework. M.F. performed the derivations. Both authors wrote and reviewed the manuscript.

\section*{Competing Interests}
The authors declare no competing interests.

\section*{Data Availability}
All data and code used in this study are available from the corresponding author upon reasonable request.

\section*{Funding}
We acknowledge support from the Knut and Alice Wallenberg Foundation (Grant No. 2023.0087), the Swedish Research Council (VR starting Grant No. 2022-03350), and the Olle Engkvist Foundation (Grant No. 229-0443).

\end{document}